\documentclass[preprint,11pt,a4]{elsarticle}

\usepackage{lineno,hyperref}
\modulolinenumbers[100]

\newcommand{\var}{\texttt}

\usepackage{float}

\usepackage{amssymb}
\usepackage{amsthm}

\usepackage{amsmath}
\usepackage{color}
\usepackage{graphicx}
\usepackage{threeparttable}
\usepackage[linesnumbered, ruled, vlined]{algorithm2e}

\usepackage{array,booktabs,ragged2e}
\newcolumntype{R}[1]{>{\RaggedLeft\arraybackslash}p{#1}}

\SetCommentSty{mycommfont}

\usepackage{amsmath}

\newtheorem{theorem}{Theorem}
\newtheorem{definition}[theorem]{Definition}

\newtheorem{proposition}[theorem]{Proposition}
\newtheorem{lemma}[theorem]{Lemma}

\usepackage{hyperref}
\usepackage{cleveref}

\usepackage[noend]{algpseudocode}
\usepackage{amsmath}
\usepackage{indentfirst}

\makeatletter
\newcommand{\rmnum}[1]{\romannumeral #1}
\newcommand{\Rmnum}[1]{\expandafter\@slowromancap\romannumeral #1@}
\makeatother

\usepackage{array}
\newcolumntype{P}[1]{>{\centering\arraybackslash}p{#1}}

\usepackage{amsmath}
\usepackage{color}
\usepackage{graphicx}
\usepackage{threeparttable}
\usepackage[linesnumbered, ruled, vlined]{algorithm2e}
\usepackage{amsmath}
\usepackage{amssymb}
\usepackage{amsthm}
\usepackage{mathtools}
\usepackage{soul}
\usepackage[normalem]{ulem}
\usepackage{times}
\usepackage{graphicx, caption, subcaption}
\usepackage{diagbox}

\usepackage{thmtools}
\usepackage{thm-restate}
\usepackage{cleveref}

\usepackage[noend]{algpseudocode}
\usepackage{indentfirst}

\usepackage{array}
\newcolumntype{C}[1]{>{\centering\arraybackslash}m{#1}}

\definecolor{azure(colorwheel)}{rgb}{0.0, 0.5, 1.0}

\definecolor{frenchblue}{rgb}{0.0, 0.45, 0.73}

\definecolor{bittersweet}{rgb}{1.0, 0.44, 0.37}

\definecolor{green(pigment)}{rgb}{0.0, 0.65, 0.31}

\definecolor{navyblue}{rgb}{0.0, 0.0, 0.5}

\definecolor{darkcerulean}{rgb}{0.03, 0.27, 0.49}

\definecolor{darkpowderblue}{rgb}{0.0, 0.2, 0.6}

\definecolor{egyptianblue}{rgb}{0.06, 0.2, 0.65}

\definecolor{caribbeangreen}{rgb}{0.0, 0.8, 0.6}

	\definecolor{darkpastelgreen}{rgb}{0.01, 0.75, 0.24}

\usepackage{float}

\usepackage{amssymb}
\usepackage{amsthm}

\usepackage{amsmath}
\usepackage{color}
\usepackage{graphicx}
\usepackage{threeparttable}
\usepackage[linesnumbered, ruled, vlined]{algorithm2e}

\makeatletter
\def\hlinewd#1{%
\noalign{\ifnum0=`}\fi\hrule \@height #1 %
\futurelet\reserved@a\@xhline}
\makeatother

\usepackage{array}
\newcolumntype{?}{!{\vrule width 1pt}}

\newtheorem{property}{Property}[section]

\usepackage[noend]{algpseudocode}
\usepackage{amsmath}
\usepackage{indentfirst}

\usepackage{array}
\newcolumntype{P}[1]{>{\centering\arraybackslash}p{#1}}

\journal{European Journal of Operational Research}

\biboptions{authoryear}

\begin{document}

\begin{frontmatter}

\title{Efficient Approximation Algorithms for Scheduling Moldable Tasks}

\author[wuxhu]{Xiaohu Wu}
\ead{xiaohu.wu@bupt.edu.cn}
\author[pl]{Patrick Loiseau}
\ead{patrick.loiseau@inria.fr}

\address[wuxhu]{Beijing University of Posts and Telecommunications, Beijing, China}
\address[pl]{Inria, FairPlay team, Palaiseau, France}





\begin{abstract}
Moldable tasks allow schedulers to determine the number of processors assigned to each task, thus enabling efficient use of large-scale parallel processing systems. We consider the problem of scheduling independent moldable tasks on processors and propose a new perspective of the existing speedup models: as the number $p$ of processors assigned to a task increases, the speedup is linear if $p$ is small and becomes sublinear after $p$ exceeds a threshold. Based on this, we propose an efficient approximation algorithm to minimize the makespan. As a by-product, we also propose an approximation algorithm to maximize the sum of values of tasks completed by a deadline; this scheduling objective is considered for moldable tasks for the first time while similar works have been done for other types of parallel tasks.
\end{abstract}


\end{frontmatter}


\section{Introduction}
\label{sec.introduction}




Most computations nowadays are done in a parallelized way on large computers containing many processors. Optimizing the use of processors leads to the problem of scheduling parallel tasks based on their characteristics. In certain cases, the number of processors assigned to a task is predefined by its owner and is said to be rigid. However, in many cases, the scheduler can decide this number before the task execution: if this number cannot be changed during the task execution, the task is said to be moldable; otherwise, it is said to be malleable\footnote{In the earlier literature, moldable tasks was also called malleable tasks. Now, malleable tasks refer to another type of parallel tasks \citep{Hanen01}.}. {Moldable tasks are easier to implement and manage than malleable tasks; the latter require additional system support for task migrations and preemptions \citep{Hanen01}.} 

\subsection{General Problem Description}


{We consider the problem of scheduling $n$ independent moldable tasks $\mathcal{T}=$ $\{T_{1}, T_{2},$ $\cdots,$ $T_{n}\}$ on $m$ identical processors;} all tasks are available at time zero. {For every task $T_{j}\in\mathcal{T}$, its execution time $t_{j,1}$ on one processor is given, as well as the speedup $\eta_{j,p}$ when assigned $p\geq 1$ processors, where $p$ is a positive integer. The execution time of $T_{j}$ on $p$ processors is $t_{j,p}=\frac{t_{j,1}}{\eta_{j,p}}$;} then, its workload is $D_{j,p}=p\times t_{j,p}$. The task $T_{j}$ can be represented by a rectangle in the processors $\times$ time space. Like \citep{Gregory99,Gregory07,Jansen18a}, given a real number $d$, we define a parameter $\gamma(j,d)$ as the minimum number of processors needed to finish task $T_{j}$ by time $d$; if $T_{j}$ cannot be finished by time $d$ on any permissible number of processors, we set by convention $\gamma(j,d)=+\infty$. We often hope to finish all tasks as soon as possible. Sometimes, a task $T_{j}$ also has a value $v_{j}$ that can be obtained if it is finished by a deadline $\tau$; then we hope to finish by time $\tau$ the most valuable tasks. We will propose algorithms that generate schedules for different objectives: (\rmnum{1}) minimize the makespan, {\em i.e.,} the maximum completion time of all tasks of $\mathcal{T}$ or (\rmnum{2}) choose a subset of tasks and finish them on the $m$ processors by a deadline $\tau$ to maximize the throughput, {\em i.e.,} the aggregate value of tasks finished by time $\tau$. For each task to be executed, a schedule will define {\em the number of processors assigned to it} and {\em the time interval in which it is finished}. An algorithm is a $\rho$-approximation if
\begin{itemize}
\item for our minimization problem, it produces a schedule whose makespan is at most $\rho$ times the optimal makespan where $\rho \geq 1$;
\item for our maximization problem, it produces a schedule whose throughput is at least $\rho$ times the optimal throughput where $\rho\leq 1$.
\end{itemize}
It is always desired to have performance bound $\rho$ closer to one, while keeping algorithms simple to run efficiently.

\subsection{Typical Speedup Models, and Motivation}

For moldable tasks, a key aspect that conditions scheduling is the relation between the task execution time $t_{j,p}$ and the number $p$ of assigned processors. Now, we introduce three typical speedup models in literature and the most related works, as well as the main motivation of this paper. {In this paper, our main problem is offline scheduling of independent moldable tasks for makespan minimization.
While introducing the related works, if they have any difference with our main problem, we only clarify their difference with ours; otherwise, they consider the same problem as our main problem.}

\begin{table}[t]
	\centering
		\caption{The Most Relevant Algorithmic Results for the Linear-speedup Model}
	\begin{threeparttable}[b]
		\begin{tabular}{|R{3.53cm}| C{3.37cm} |  C{4.2cm} |}
			\hline
                            &    {Optimality or Approximation} Ratio              &   Remarks       \\ \hline
          \cite{Wang92a}    &   $3-\frac{2}{m}$                          &   Dependent     \\ \hline
    \cite{Drozdowski96a}    &     {Exact}                    &   Malleable, {Polynomial time solvable}      \\ \hline	
          \cite{Jain15a}    &   $\frac{m-k}{m}\frac{s-1}{s}$             &   Malleable, Throughput maximization      \\ \hline	
             \cite{Jain}    &   $2+$$\mathcal{O}\left(\frac{1}{(\sqrt[3]{s}-1)^{2}}\right)$   &  Malleable, Online, Throughput maximization       \\ \hline
            \cite{Wu15a}    &   $\frac{s-1}{s}$ \& exact respectively   &  Malleable, Throughput maximization       \\ \hline
 \cite{guo2017efficient}    &   $\frac{m-k}{m}$ \& exact respectively   &  Malleable, Throughput maximization       \\ \hline
        \cite{Benoit20a}    &   2                 &   for failure-prone platforms     \\ \hline
 \cite{benoit2022online}    &   2.62              &   Dependent, Online     \\ \hline
		\end{tabular}
	\end{threeparttable}
	\label{table-linear-model}
\end{table}

\vspace{0.18em}\noindent\textbf{Linear-Speedup Model.} {An ideal speedup model is linear when $p$ does not exceed a threshold $\delta_{j}$ \citep{Hanen01}: $t_{j,p}=\frac{t_{j,1}}{p}$ where $\eta_{j,p}=p$; the workload of $T_{j}$ is independent of $p$ since $D_{j,p}=pt_{j,p}=t_{j,1}$.} \cite{Benoit20a} propose a 2-approximation algorithm, called LPA-LIST, for failure-prone platforms with additional constraints in the process of executing jobs. {We note that LPA-LIST is applicable to the main problem of this paper by setting the number of job execution failures in its model to zero. Like ours, the other related works of this paper are directly for failure-free platforms.} When there are precedence constraints among moldable tasks, \cite{Wang92a} propose a $\left(3-\frac{2}{m}\right)$-approximation algorithm {while \cite{benoit2022online} give a 2.62-approximation algorithm in the online setting.} Besides, the case of scheduling independent malleable tasks has already been studied well, e.g., {\cite{Drozdowski96a}} {gives} {a polynomial time exact algorithm with a time complexity of $\mathcal{O}(n^{2})$.} {Table \ref{table-linear-model} summarizes the most relevant works under this model and their differences with our main problem are clarified in the third column; here, the works whose objectives are throughput maximization will be introduced in Section \ref{sec.throughput-maximization}.}


\begin{table}[t]
	\centering
		\caption{Algorithmic Results for the Communication {Time} Model}
	\begin{threeparttable}[b]
		\begin{tabular}{|R{3.4cm}| C{4.9cm} |  C{3.1cm} |}
			\hline
                            &      {Approximation} Ratio                                               &   Remarks       \\ \hline
        \cite{Dutton07a}    &   $\frac{30}{13}$ when $m\rightarrow \infty$; 2, $\frac{9}{4}$ and $\frac{20}{9}$ for $m = 2$, 3 and 4 respectively  &   Online, $c_{j}=c$     \\ \hline
        \cite{Havill08a}    &   $4 - \frac{4}{m}$ for even $m\geq 2$; $4-\frac{4}{m+1}$ for odd $m\geq 3$        &   Online, $c_{j}=c$     \\ \hline
    \cite{guo2010online}    &   $\frac{1+\sqrt{5}}{2}$ for $m=2$                            &   Online   \\ \hline
 \cite{kell2015improved}    &   1.5 for $m= 2$; 2 for $m=3$                           &   Online, $c_{j}=c$      \\ \hline	
        \cite{Benoit20a}    &   3                                                     &   For failure-prone platforms     \\ \hline
 \cite{benoit2022online}    &   3.61                                                  &   Dependent, Online     \\ \hline
		\end{tabular}
	\end{threeparttable}
	\label{table-communication-model}
\end{table}

\vspace{0.18em}\noindent\textbf{Communication {Time} Model.} {The communication {time} model is defined by a function:
\begin{align}\label{equa-communication-model}
t_{j,p} = \frac{t_{j,1}}{p} + (p-1)c_{j},
\end{align}
where $c_{j}$ is a positive real number; the term $(p-1)c_{j}$ is used to model the communication overhead among different parts of a task. As more processors are assigned, the overhead and workload $D_{j,p}$ increase; if $p$ is too large, $t_{j,p}$ will not decrease and even increase as $p$ increases, due to the effect of $(p-1)c_{j}$.} {Like Table \ref{table-linear-model}, Table \ref{table-communication-model} summarizes the related works.} {Specifically, when all tasks $T_{j}\in\mathcal{T}$ have the same $c_{j}=c$, \cite{Dutton07a} give an online algorithm whose approximation ratio is 2, $\frac{9}{4}$, and $\frac{20}{9}$ for $m = 2$, 3, and 4 respectively, and is $\frac{30}{13}$ when $m\rightarrow \infty$}. \cite{Havill08a} propose an online algorithm with an approximation ratio $\frac{4(m-1)}{m}$ for even $m\geq 2$ and $\frac{4m}{m+1}$ for odd $m\geq 3$. {\cite{kell2015improved} improve the work of \citep{Dutton07a} by giving online algorithms whose approximation ratio are 1.5 and 2 for $m = 2$ and 3.} The following works consider the case that each task $T_{j}$ has a specific $c_{j}$. \cite{guo2010online} give an online algorithm whose approximation ratio is $(1+\sqrt{5})/2$ for $m=2$, and show that $(1+\sqrt{5})/2$ is a lower bound on the approximation ratio of any online algorithm for the problem with $m\geq 2$. In the offline setting, \cite{Benoit20a} show that LPA-LIST is a 3-approximation for failure-prone platforms. {\cite{benoit2022online} consider online scheduling of moldable tasks with precedence constraints and give a 3.61-approximation algorithm.} 

\begin{figure*}[t]
  \begin{center}
  \includegraphics[width=4.5in]{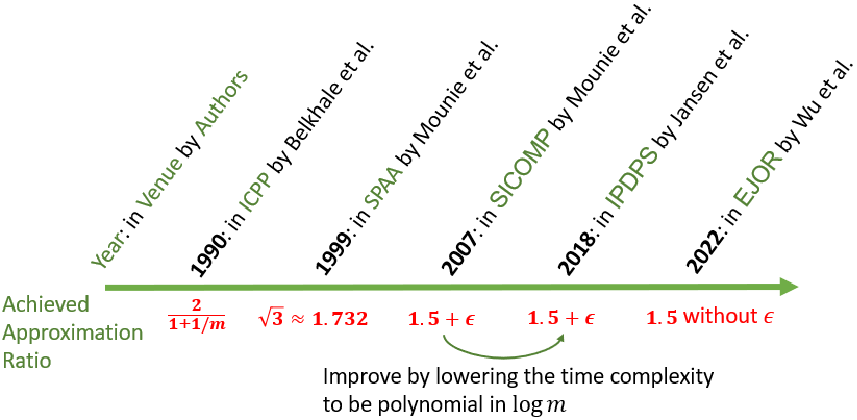}
  \end{center}
  \caption{Major Algorithmic Improvements for the Monotonic Model over the Past Three Decades.}\label{Fig-monotonic-progress}
\end{figure*}

\vspace{0.18em}\noindent\textbf{Monotonic Model.} {To date, the best algorithm for our problem is designed by simply using a general monotonic assumption: $t_{j,p}$ is non-increasing and $D_{j,p}$ is non-decreasing in $p\in [1, m]$, where $\eta_{j,p}\leq p$.} {Fig. \ref{Fig-monotonic-progress} illustrates the major algorithm improvements over the past three decades, where $m$ is independent of $n$.} Specifically, \cite{Belkhale90a} give a $\frac{2}{1+1/m}$-approximation algorithm. \cite{Gregory99,Gregory07} first propose a ($\sqrt{3}+$ $\epsilon$)-approximation algorithm and then a ($\frac{3}{2}+\epsilon$)-approximation algorithm with a complexity $\mathcal{O}(mn\log{\frac{n}{\epsilon}})$ where $\epsilon$ is arbitrarily small. \cite{Jansen18a} achieve an improved complexity polynomial in $\log{m}$ and $\frac{1}{\epsilon}$ and linear in $n$, although the algorithm is still a ($\frac{3}{2}+\epsilon$)-approximation. Additionally, in the special case where $m\geq 8\frac{n}{\epsilon}$, they give a FPTAS with a complexity $\mathcal{O}( n\log^{2}{m}( \log{m} + \log{\frac{1}{\epsilon}} ) )$. The FPTAS requires a specific relation between $n$ and $m$. {\cite{wu2022improved} give a $\frac{3}{2}$-approximation algorithm without $\epsilon$ and its time complexity is $\mathcal{O}(mn\log(mn))$ for $m>n$ and $\mathcal{O}(n^{2}\log n)$ for $m\leq n$.} 
{As illustrated in Fig. \ref{Fig-monotonic-progress}, the three recent algorithmic results all have approximation ratios of around 1.5, and it is difficult to lower the best known approximation ratio 1.5. In this paper, we aim to sacrifice the generality of the monotonic model for a better performance guarantee.
}

In the case where $n$ is independent of $m$, we hope to develop a $\rho$-approximation algorithm with $\rho<\frac{3}{2}$ and will revisit the related speedup models. Under the monotonic assumption, we have the following bounds of the execution time $t_{j, \gamma(j, d)}$ when a task $T_{j}$ is assigned $\gamma(j,d)$ processors, which will also hold in this paper:
\begin{equation}\label{monotonic-inequality}
d\geq t_{j, \gamma(j, d)} > d(\gamma(j,d)-1)/\gamma(j,d).
\end{equation}
By the definition of $\gamma(j,d)$, $D_{j,\gamma(j,d)}$ is the minimum workload needed to complete $T_{j}$ by time $d$. Suppose that an algorithm produces a schedule of a makespan $d$. We observe that it is a $\frac{1}{\theta}$-approximation to makespan minimization if every task $T_{j}\in$ $\mathcal{T}$ has the minimum workload and the aggregate workload processed on the $m$ processors in $[0, d]$ is $\geq \theta md$ where $\theta$ is a lower bound of the processor utilization. Our objective is to make $\theta$ large (e.g., $\theta > \frac{2}{3}$). For each task $T_{j}$ with large $\gamma(j,d)$ (e.g., $\gamma(j,d)\geq 4$), we have by Inequality (\ref{monotonic-inequality}) that executing it on $\gamma(j,d)$ processors alone can make these processors achieve a high utilization in $[0, d]$. One main challenge comes from tasks with smaller $\gamma(j,d)$. Then, a more precise speedup description than monotonicity could help, which is fortunately available in literature; it allows quantitatively characterizing the execution time reduction while keeping the workload constant, when the number $p$ of processors assigned to a task $T_{j}$ changes from $\gamma(j,d)$ to a larger value. We can thus obtain some desired properties and design a schedule under which the $m$ processors achieve a high overall utilization in $[0, d]$ under some additional constraints (see Section~\ref{sec.high-idea}).

\vspace{0.18em}\noindent\textbf{The Proposed Speedup Model.} {While the linear-speedup model is studied, \citep{Drozdowski96a} points out that it is typical of parallel applications that the speedup is linear when $p$ is within a relatively small $\delta_{j}$; assigning more than $\delta_{j}$ processors to execute $T_{j}$ becomes less efficient. This model sets the parallelism bound of $T_{j}$ to be $\delta_{j}$, although it may be worth exploring the opportunity of assigning more processors to each task $T_{j}$ to get better resource efficiency. Complementarily, the function (\ref{equa-communication-model}) of the communication {time} model is also tested on widely used NAS parallel benchmarks and HPLinpack, which embody various computations with typical communication patterns for evaluating the performance of parallel systems \citep{John18a}; here, an instance of a type of computation represents a task. The benchmarking results of \cite{Dutton08} show that the function (\ref{equa-communication-model}) can well approximate the execution times of tasks and also indicate that the factor $c_{j}$ is far smaller than $t_{j,1}$: when $p$ is small (up to a threshold $\delta_{j}$), the effect of $(p-1)c_{j}$ on $t_{j,p}$ is negligible compared with the term $\frac{t_{j,1}}{p}$ and the speedup coincides accurately with the linear-speedup model \citep{Drozdowski96a}; assigning more than $\delta_{j}$ processors to execute $T_{j}$ becomes less efficient: its execution time still decreases as $p$ increases but its workload starts to increase, similarly to monotonic tasks; finally, there may be a larger threshold $k_{j}$ such that when $p>k_{j}$, its execution time does not decrease any longer and even increases as $p$ increases, since parallelizing on too many processors incurs an unacceptable overhead.} Thus, we associate every task $T_{j}$ with two thresholds $\delta_{j}$ and $k_{j}$ to distinguish the speedup modes of $T_{j}$ when $p$ is in different ranges where $\delta_{j}\leq k_{j}$; then, we make the following definition on which we will base the algorithmic design of this paper.

\begin{definition}\label{def.deltak-monotonic-task}
{A task $T_{j}\in\mathcal{T}$ is $(\delta_{j}, k_{j})$-monotonic if it is moldable and satisfies}
\begin{enumerate}
\item {When $p\in [1, \delta_{j}]$, its workload remains constant and the speedup is linear}, {\em i.e.,} $D_{j,1}=D_{j,p}=p\times t_{j,p}$;

\item If $\delta_{j}<k_{j}$, its workload is increasing and its execution time is decreasing in $p\in [\delta_{j}, k_{j}]$, i.e., $D_{j,p}< D_{j,p+1}$ and $t_{j,p}>t_{j,p+1}$ for $p\in [\delta_{j}, k_{j}-1]$.

\item {The parameter $k_{j}$ is a parallelism bound, {\em i.e.,} the maximum number of processors allowed to be assigned to $T_{j}$.}
\end{enumerate}
\end{definition}

\begin{figure*}[t]
  \begin{center}
  \includegraphics[width=4.45in]{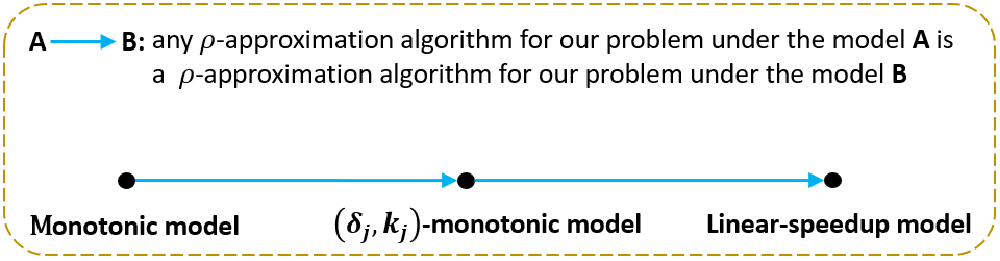}
  \end{center}
  \caption{Relations Between Different Speedup Models}\label{Fig-Generalization-Level}
\end{figure*}

In Definition~\ref{def.deltak-monotonic-task}, the second point implies that assigning more than $\delta_{j}$ for executing $T_{j}$ is less efficient but its execution time is decreasing in $p\in [\delta_{j}, k_{j}]$. The third point is used to reflect that when $p>k_{j}$, the workload begins to increase to an unacceptable extent such that the execution time does not decrease any more (i.e., $\eta_{j,k_{j}}\geq \eta_{j,p}$); then, assigning more than $k_{j}$ processors to $T_{j}$ cannot bring any benefit. {Overall, when $p\in [1, k_{j}]$, $t_{j,p}$ is non-increasing in $p$ while $D_{j,p}$ is non-decreasing in $p$.

\vspace{0.12em}\noindent\textbf{Relations with the Monotonic and Linear-Speedup Models.} {For each task $T_{j}\in\mathcal{T}$, the speedup model defines the way that $t_{j,p}$ and $D_{j,p}$ change with the number $p$ of allocated processors, where $t_{j,1}$ is known and $D_{j,p}=p\times t_{j,p}$. We consider the problem of offline scheduling of independent moldable tasks on identical machines, and the objective is either makespan minimization or throughput maximization. By Definition~\ref{def.deltak-monotonic-task}, a task whose speedup is linear is also a $(\delta_{j}, k_{j})$-monotonic task when $k_{j}=\delta_{j}$. Thus, the linear-speedup model is a special case of the $(\delta_{j}, k_{j})$-monotonic model; thus, for a given objective, any $\rho$-approximation algorithm for the problem under the $(\delta_{j}, k_{j})$-monotonic model of this paper is also a $\rho$-approximation algorithm for the problem under the linear-speedup model, as illustrated in Fig. \ref{Fig-Generalization-Level}. A problem $A$ is S-reducible to a problem $B$ if any instance of $A$ can be transformed into an instance of $B$ with the same optimal objective function value, and any solution for $B$ can be transformed into a solution for $A$ with the same objective function value \citep{crescenzi2016complexity}. There exists a S-reduction from the problem under the $(\delta_{j}, k_{j})$-monotonic model to the problem under the monotonic model, which is proved in \ref{appendix-monotonic}, and thus any $\rho$-approximation algorithm for the monotonic model can be transformed into a $\rho$-approximation algorithm for the $(\delta_{j}, k_{j})$-monotonic model.

Algorithms for scheduling problems with a more general speedup model have more extensive applicability, as illustrated in Fig. \ref{Fig-Generalization-Level}. However, algorithms under specific models are still important since they may be designed more finely to have better approximation ratios. For example, for online scheduling of moldable task graphs to minimize the makespan, \citep{benoit2022online} give a 2.62-approximation algorithm for the linear-speedup model and a 3.61-approximation algorithm for the communication {time} model; they also generalize these speedup models and give a 5.72-approximation algorithm under the generalized model.

\subsection{Algorithmic Results}

Consider a set $\mathcal{T}$ in which each task $T_{j}$ is $(\delta_{j}, k_{j})$-monotonic. Given a task $T_{j}$, its parameters $\delta_{j}$ and $k_{j}$ are fixed; as reported in \citep{Dutton08}, $\delta_{j}$ and $k_{j}$ typically range in $[25, 150]$ and $[250, 512]$, depending on the types of computation embodied in the tasks of $\mathcal{T}$. {We denote by $\delta$ the minimum linear-speedup threshold of all tasks} and by $k$ the maximum parallelism bound of all tasks, i.e.,
\begin{align}\label{equa-min-delta-max-k}
{\delta=\min\nolimits_{T_{j}\in\mathcal{T}}\{\delta_{j}\} \text{ and } k=\max\nolimits_{T_{j}\in\mathcal{T}}\{k_{j}\}.}
\end{align}
The number $m$ of processors is large since our problem arises in large-scale parallel systems such as supercomputers and cloud computing clusters \citep{Jain15a,Aridor05a}, {\em e.g.,} supercomputers can have $m=2^{16}$ processors inside \citep{Aridor05a}. Like \citep{Jain15a}, we assume in this paper that $m$ is much larger than the maximum parallelism bound of tasks, i.e., $m\gg k.$

{Let $u=\left\lceil \sqrt[2]{\delta} \right\rceil-1$, that is, $u$ is the unique integer such that $\delta\in [u^{2}+1, (u+1)^{2}]$. {Let $t_{m}$ denote the maximum execution time of tasks when they are executed on one processor, i.e., $t_{m}=\max_{T_{j}\in\mathcal{T}}\{t_{j,1}\}$.} In this paper, for any $\delta\geq 5$, {\em the main algorithmic result} is a $\frac{1}{\theta(\delta)} (1+\epsilon)$-approximation algorithm for makespan minimization with a complexity {$\mathcal{O}(n\log{m}\log{(n m t_{m}/\epsilon)})$} where
    \begin{align*}
     \theta(\delta) = \frac{u+1}{u+2}\left( 1 -  \frac{k}{m} \right).
    \end{align*}
The algorithm achieves an approximation ratio $\theta(\delta)$ close to $\frac{u+2}{u+1}$ since $m\gg k$. Typically, the minimum linear-speedup threshold $\delta$ has an effective range of $[25, 150]$ \citep{Dutton08}. In the worst case that $\delta=25$, $\theta(\delta)$ is close to $\frac{6}{5}$; when $\delta=150$, $\frac{u+2}{u+1}=\frac{14}{13}\approx 1.077$, which is close to 1. The larger the threshold $\delta$, the better the proposed algorithm. Under mild assumptions, we realize our goal to sacrifice the generality of the monotonic model for a better approximation ratio.}


For throughput maximization with a given deadline $\tau$, we assume that every task $T_{j}\in\mathcal{T}$ can be finished by time $\tau$, i.e., {$\gamma(j,\tau)\in [1, k_{j}]$}. As a by-product, \textit{another algorithmic result of this paper} is a $\theta(\delta)$-approximation algorithm with a complexity {$\mathcal{O}(n^{2}\log{m})$} to maximize the throughput with a deadline $\tau$. To the best of our knowledge, we are the first to address this scheduling objective for moldable tasks, while this objective has been addressed for other types of parallel tasks in the literature of scheduling theory \citep{Jansen04,Jansen05a}.

The rest of this paper is organized as follows. In Section~\ref{sec.related-work}, we give more related works. In Section~\ref{sec.high-idea}, we
give an overview of the ideas developed in this paper. The following two sections are used to elaborate these ideas. In particular, in Section~\ref{sec.utilization}, we propose a scheduling algorithm $Sched$ that produces a schedule with several features described in Section~\ref{sec.high-idea}. In Section~\ref{sec.objectives}, we show the application of $Sched$ to the objectives of makespan minimization and throughput maximization with a deadline respectively. Finally, we conclude this paper in Section~\ref{sec.conclusion}.

\section{Related Work}
\label{sec.related-work}

\subsection{Makespan Minimization}

The problem of scheduling moldable tasks to minimize the makespan is strongly NP-hard when $m\geq 5$ \citep{Hanen01}. There is a long history of study with continuous improvements to the approximation ratio or time complexity. \cite{Turek92a} consider moldable tasks without monotonicity and propose a two-phases approach: (\rmnum{1}) determine the number of processors assigned to each task and (\rmnum{2}) solve the resulting strip packing problem; the latter has well been studied, {\em e.g.,} we can directly use the 2-approximation algorithm of Steinberg \citep{Steinberg97a}. Further, the authors show that any $\lambda$-approximation algorithm of a complexity $\mathcal{O}(f(m,n))$ for strip packing can be transformed into a $\lambda$-approximation algorithm of a complexity $\mathcal{O}(mnf(m,n))$ for our problem. In the special case of monotonic tasks, \cite{Ludwig94a} improve the transformation complexity to $\mathcal{O}(n\log^{2}{m}+f(m,n))$. \cite{Jansen99} formulate the original problem as a linear program. They propose a polynomial time approximation scheme (PTAS) when the number $m$ of processors is constant; here, the complexity is exponential in $m$. Further, \cite{Jansen10a} propose a PTAS when $m$ is polynomially bounded in the number $n$ of tasks. In the case of an arbitrary number of processors, \cite{Jansen12a} also propose a polynomial time ($\frac{3}{2}+\epsilon$)-approximation algorithm for any fixed $\epsilon$. {\cite{barketau2014scheduling} give an optimal enumerative algorithm whose time complexity is $\mathcal{O}(n^{3}2^{(2n+m-2)n})$.} In the special case of $n$ identical tasks, \cite{Decker06a} give a $\frac{5}{4}$-approximation algorithm.

As introduced in Section~\ref{sec.introduction}, of the great relevance to our work are \citep{Gregory99,Gregory07,Jansen18a} that use similar techniques for monotonic tasks. For example, \cite{Gregory07} apply the dual approximation technique \citep{Hochbaum87a}: it takes a real number $d$ as an input, and either outputs a schedule of a makespan $\leq \frac{3}{2}d$ or answers correctly that $d$ is a lower bound of the optimal makespan. To realize this, tasks are mainly classified into two subsets $\mathcal{T}_{1}$ and $\mathcal{T}_{2}$ whose tasks are respectively assigned $\gamma(j,d)$ and $\gamma(j,\frac{d}{2})$ processors; the classification aims at minimizing the total workload $W$ of $\mathcal{T}_{1}$ and $\mathcal{T}_{2}$ while guaranteeing that the total number of processors assigned to $\mathcal{T}_{1}$ is $\leq m$, which is formulated as a knapsack problem. If the optimal $W$ exceeds the processing capacity of the $m$ processors, there exists no schedule with a makespan $<d$. Otherwise, the total number of processors assigned to $\mathcal{T}_{2}$ may exceed $m$ and a series of reductions to the numbers of processors assigned to the tasks of $\mathcal{T}_{1}$ and $\mathcal{T}_{2}$ is taken to get a feasible schedule: the tasks are assigned to different parts of processors respectively in the time intervals $[0, \frac{3}{2}d]$, $[0, d]$ and $[d, \frac{3}{2}d]$.

{Finally, our problem has also been studied well when the speedup $\eta_{j,p}$ is a concave or convex function of $p$ \citep{blazewicz2004scheduling,1608009,barketau2014scheduling,ebrahimi2018scheduling}, which is less relevant to the speedup model of this paper. We don't introduce them in this paper any more.}

\subsection{Throughput Maximization}
\label{sec.throughput-maximization}


Several works have considered scheduling other types of parallel tasks to maximize the throughput. \cite{Jansen04} and \cite{Jansen05a} consider scheduling rigid tasks with a common deadline, {\em e.g.,} the former apply the theory of knapsack problem and linear programming to propose an ($\frac{1}{2}+\epsilon$)-approximation algorithm. \cite{Jain15a} consider malleable tasks with individual deadlines. Each task has a linear speedup within a parallelism bound, and there is a parameter $s$ used to characterize the minimum delay-tolerance of all tasks: each $T_{j}$ has to be finished in a time window $[a_{j}, d_{j}]$; it has the minimum execution time $len_{j}$ when assigned $\delta_{j}$ processors; $s$ is the minimum ratio of $d_{j}-a_{j}$ to $len_{j}$ among all tasks. For offline scheduling, \cite{Jain15a} propose a greedy $\frac{m-k}{m}\frac{s-1}{s}$-approximation algorithm where $k$ is the maximum parallelism bound of all tasks. \cite{Wu15a} prove that the best approximation ratio that the type of greedy algorithms of \citep{Jain15a} can achieve is $\frac{s-1}{s}$ and propose such an algorithm with a time complexity of $\mathcal{O}(n^{2})$; they also show a sufficient and necessary condition under which a set of malleable tasks with deadlines can feasibly be scheduled on a fixed number of processors and propose an exact algorithm by dynamic programming that has a time complexity of $\mathcal{O}\left( \max\{n^{2}, n(mT)^{T}\} \right)$, where $T$ is the maximum deadline of tasks. {\cite{guo2017efficient} give a $\frac{m-k}{m}$-approximation algorithm with a time complexity of $\mathcal{O}(n^{2}+nT)$ and also an exact algorithm with a time complexity of $\mathcal{O}\left( n(mT)^{T} \right)$.} For online scheduling, \cite{Jain} propose a $2+$$\mathcal{O}\left(1/(\sqrt[3]{s}-1)^{2}\right)$-approximation algorithm. In cloud computing clusters, many applications are delay-tolerant where $s\gg 1$ and $m\gg k$. Thus, their algorithms achieve good approximation ratios in practical settings.

\section{Overview of the Approaches}
\label{sec.high-idea}

Central to our algorithm design is an algorithm $Sched$ that aims to schedule a set $\mathcal{T}$ of tasks on the $m$ processors in a time interval $[0, d]$ and achieves a processor utilization $\geq\theta(\delta)$ on the conditions that (\rmnum{1}) each scheduled task $T_{j}$ has a workload $D_{j,\gamma(j,d)}$, which is the minimum workload to finish $T_{j}$ by time $d$, and (\rmnum{2}) there exists some task of $\mathcal{T}$ rejected to be scheduled due to the insufficiency of idle processors (see Section~\ref{sec.utilization}). We establish the connection of $Sched$ with our two problems in the following ways.

For makespan minimization, we need to schedule all tasks of $\mathcal{T}$, while $Sched$ can play a role only when a part of tasks are scheduled. We apply a binary search procedure to find two parameters $U$ and $L$ such that $Sched$ can schedule all tasks by time $U$ but only a part of tasks by time $L$, with the relation $U$ $\leq$ $L (1+\epsilon)$ ({see Section~\ref{sec.makespan}}). Let $d^{*}$ denote the optimal makespan. We can establish the relation between $U$ and $d^{*}$ via $L$ and prove $U/d^{\ast} \leq \frac{1}{\theta(\delta)}(1+\epsilon)$, thus showing that the resulting algorithm is a $\frac{1}{\theta(\delta)}(1+\epsilon)$-approximation. Specifically, in the case that $d^{*}\in [L, U]$, we have $U/d^{\ast}\leq (1+\epsilon)/\theta(\delta)$ trivially. In the case that $d^{*}< L$, we have that the total workload of all tasks of $\mathcal{T}$ in an optimal schedule is $\leq md^{*}$ but $\geq$ the total workload processed when $Sched$ manages to schedule a part of tasks of $\mathcal{T}$ by time $L$. Thus, we have $md^{*}\geq m\theta(\delta) L\geq m\theta(\delta) U/(1+\epsilon)$ and $U$ $\leq$ $d^{*} (1+\epsilon)/\theta(\delta)$.

For throughput maximization, $v_{j}/D_{j, \gamma(j,d)}$ is the maximum possible value obtained from processing a unit of workload of $T_{j}$, called its value density. Let us accept the maximum number of tasks in the non-increasing order of their value densities until $Sched$ cannot produce a feasible schedule by time $\tau$; then, the feature of $Sched$ leads to that the utilization $\theta(\delta)$ will be the approximation ratio of the resulting algorithm ({see Section~\ref{sec.model2}}).

Finally, the design of $Sched$ relies on the properties of the speedup model in Definition~\ref{def.deltak-monotonic-task} to classify the tasks of $\mathcal{T}$. The threshold $\delta$ in Equation (\ref{equa-min-delta-max-k}) is a fixed parameter and we have the following property by Definition~\ref{def.deltak-monotonic-task}.
\begin{property}\label{property-deltak-tasks}
If a task $T_{j}$ is ($\delta_{j}$, $k_{j}$)-monotonic, we have that (\rmnum{1}) the workload $D_{j,p}$ is non-decreasing and the execution time $t_{j,p}$ is non-increasing in the number $p$ of assigned processors when $p\in [1, k_{j}]$ and (\rmnum{2}) the speedup is linear when $p\in [1, \delta]$, i.e., $t_{j,p}=\frac{t_{j,1}}{p}$.
\end{property}

For a task $T_{j}\in\mathcal{T}$, its execution time on $p$ processors is defined by $t_{j,1}$ and $\eta_{j,p}$. Given the time $d$, $\gamma(j,d)=\min\{p\in [1, k_{j}]\, |\, t_{j,p}\leq d\}$ is a fixed parameter and can be found by binary search \citep{Jansen18a}. {The classification of tasks for the scheduling process} mainly uses three integer variables $\nu$, $H$ and $\delta^{\prime}$ and is based on the values of $\gamma(j,d)$, $t_{j,\gamma(j,d)}$ and $t_{j,\delta^{\prime}}$; it attempts to guarantee that the aggregate execution time is in $[rd, d]$ when some tasks in the same class are executed on a group of $\gamma(j,d)$ or $\delta^{\prime}$ processors. Specifically, $\nu$ and $H$ are for distinguishing tasks with different $\gamma(j,d)$: a task $T_{j}$ is said to have a large, medium, or small $\gamma(j,d)$ if $\gamma(j,d)$ is $\geq H$, in $[\nu,$ $H-1]$, or $\leq \nu-1$ {respectively}, where $\nu < H$. Let $r=\frac{H-1}{H}$ and we will use $rd$ and $(1-r)d$ to distinguish tasks with different execution times. The first class of tasks, denoted by $\mathcal{A}^{\prime}$, includes every task that has a large execution time $\geq rd$ when assigned a group of $\gamma(j,d)$ processors (see Equation (\ref{equa-first-class})), e.g., every task with large $\gamma(j,d)$ has such a feature by Inequality (\ref{monotonic-inequality}).

For the remaining tasks with medium or small $\gamma(j,d)$, we will maintain several relations among $\nu$, $H$, $\delta^{\prime}$ and $\delta$. For example, by letting $H-1\leq \delta^{\prime}\leq \delta$, the speedup is linear and the workload keeps constant when the number $p$ of assigned processors ranges in $[\gamma(j,d), \delta^{\prime}]$. These relations finally enable the following properties:
\begin{itemize}
\item For the tasks with small $\gamma(j,d)$ whose execution times are $<rd$ when assigned $\gamma(j,d)$ processors, they are denoted by $\mathcal{B}_{\nu-1}$ and their execution times will decrease remarkably (by a factor at least $\frac{\delta^{\prime}}{\nu-1}$) to a small value $<(1-r) d$ when assigned $\delta^{\prime}$ processors (see Equation (\ref{equa-B-nu-1}) and Lemma~\ref{lemma-B-nu-1}). Executing as many such tasks as possible on a group of $\delta^{\prime}$ processors in $[0, d]$ will lead to an aggregate execution time $\geq r d$.
\item Let $h$ be an integer in $[\nu, H-1]$. For the tasks with $\gamma(j,d)=h$ whose execution times are $\geq (1-r)d$ when assigned $\delta^{\prime}$ processors and $< rd$ when assigned $h$ processors, they are denoted by $\mathcal{A}_{h}$ and there exists a positive integer $x_{h}$ such that the aggregate execution time is in $[rd, d]$ when $x_{h}$ such tasks are executed one by one on a group of $\delta^{\prime}$ processors (see Equation (\ref{equa-A-h}) and Proposition~\ref{lemma-x-h}).
\end{itemize}
Finally, each group of $\gamma(j,d)$ or $\delta^{\prime}$ assigned processors described above can achieve a utilization $\geq r$ in $[0, d]$. The overall utilization $\theta(\delta)$ of the $m$ processors is close to $r$ and can be derived when some task is rejected due to the insufficiency of processors, with at most $k-1$ processors idle. The task classification and maintained relations are formally described in Section~\ref{sec.task-classification}, with other related issues solved. The scheduling algorithm $Sched$ is given in Section~\ref{sec.algo-definition}.

\section{The Algorithm $Sched$}
\label{sec.utilization}

In this section, we consider the case that every task $T_{j}\in\mathcal{T}$ can be finished by time $d$, i.e., $\gamma(j,d)\in [1, k_{j}]$.

\begin{lemma}\label{lemma-monotonic-property}
For every $(\delta_{j}, k_{j})$-monotonic task $T_{j}\in \mathcal{T}$, Inequality (\ref{monotonic-inequality}) holds.
\end{lemma}
\begin{proof}
We have $t_{j,\gamma(j,d)}\leq d$ and $t_{j,\gamma(j,d)-1}>d$ by the definition of $\gamma(j,d)$. By Property~\ref{property-deltak-tasks}, $D_{j,\gamma(j,d)}\geq D_{j,\gamma(j,d)-1}$. Further, we have $\gamma(j,d) t_{j,\gamma(j,d)} \geq (\gamma(j,d)-1) t_{j,\gamma(j,d)-1} > (\gamma(j,d)-1) d$. Hence, Inequality~\eqref{monotonic-inequality} holds.
\end{proof}

\subsection{Task Classification}
\label{sec.task-classification}

Following the high-level ideas in Section~\ref{sec.high-idea}, we now begin to elaborate the task classification. For ease of reference, we first summarize the maintained relations {between the fixed parameter $\delta$, the integer variables $H$, $\nu$, $\delta^{\prime}$, $x_{\nu}$, $\cdots$, $x_{H-1}$, and the number $r=\frac{H-1}{H}$ where the meanings of these variables and the number $r$ will be clarified later}:
\begin{subequations}
\begin{align}
    & 1\leq  \nu  \leq H  -1 \leq \delta^{\prime} \leq \delta\label{bound-1b}\\
    & \frac{r \nu}{\delta^{\prime}} \geq 1 - r\label{bound-1}\\
    & \frac{r (\nu- 1)}{\delta^{\prime}}  < 1-r\label{bound-2}
\end{align}
\end{subequations}
and for all $h\in [\nu, H-1]$
\begin{subequations}
\begin{align}
    & r \frac{h}{\delta^{\prime}} x_{h} \leq 1, \label{bound}\\
    & \max\left\{1-r, \frac{h-1}{\delta^{\prime}} \right\} x_{h} \geq r.\label{boundb}
\end{align}
\end{subequations}
As we classify tasks and prove their properties, we can gradually perceive the underlying reasons why these relations are established to get the desired properties. At the end of this subsection, we will give a feasible solution of $H$, $\nu$, $\delta^{\prime}$, $x_{\nu}$, $\cdots$, $x_{H-1}$ that satisfy the relations~(\ref{bound-1b})-(\ref{boundb}).

\begin{figure*}[t]
  \begin{center}
  \includegraphics[width=5.0in]{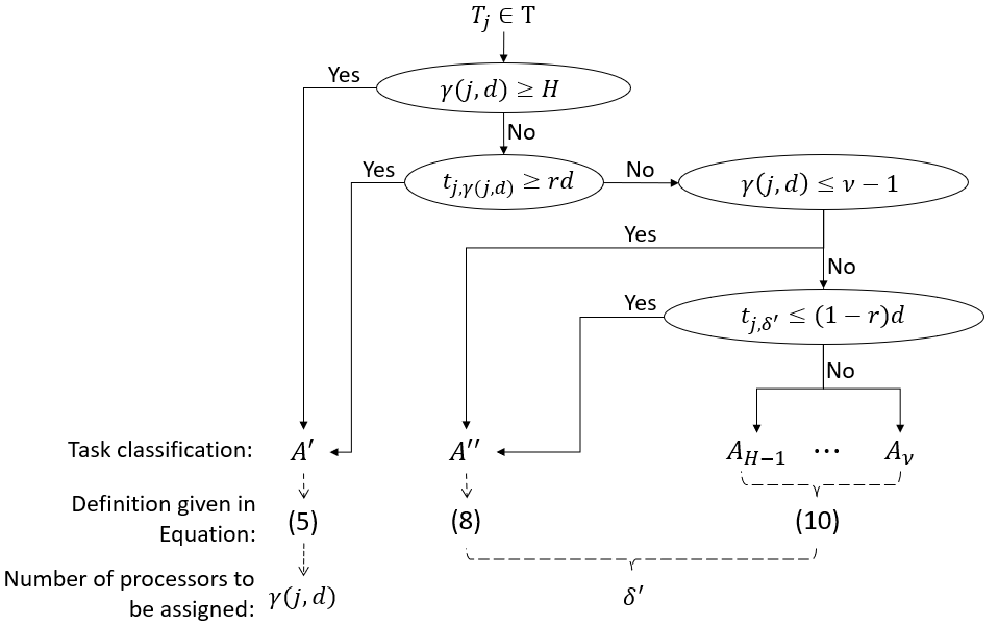}
  \end{center}
  \caption{\footnotesize{Task Classification.}}\label{Fig-Classification}
\end{figure*}

Fig.~\ref{Fig-Classification} summarizes how to classify a task $T_{j}\in\mathcal{T}$ according to its value of $\gamma(j,d)$ and its execution time on $\gamma(j,d)$ or $\delta^{\prime}$ processors. Specifically, {\em the first class of tasks} contains all tasks whose execution times $t_{j,\gamma(j,d)}$ are $\geq r d$ when assigned $\gamma(j,d)$ processors and is defined as
\begin{equation}\label{equa-first-class}
\begin{split}
\boldsymbol{\mathcal{A}^{\prime}} = \{ & T_{j}\in\mathcal{T} | \gamma(j,d)\geq H \} \\
& \cup \left\{ T_{j}\in\mathcal{T} | \gamma(j,d)\in [1, H-1], t_{j, \gamma(j,d)}\geq r d \right\} 
\end{split}
\end{equation}
$\mathcal{A}^{\prime}$ also includes a part of tasks with smaller $\gamma(j,d)$ but they have $t_{j,\gamma(j,d)}\geq r d$. 
Except $\mathcal{A}^{\prime}$, the remaining tasks have medium or small $\gamma(j,d)$ and each has an execution time $t_{j,\gamma(j,d)}<rd$. Among these tasks, let $\mathcal{B}_{\nu-1}$ denote all tasks with $\gamma(j,d)\leq \nu-1$, i.e.,
\begin{align}\label{equa-B-nu-1}
\mathcal{B}_{\nu-1} = \{T_{j}\in\mathcal{T}\, |\, \gamma(j,d)\leq \nu-1, t_{j,\gamma(j,d)} < r d \};
\end{align}
let $\mathcal{B}_{H-1}$ denote all tasks that satisfy $\gamma(j,d)\in [\nu, H-1]$ and $t_{j,\delta^{\prime}}< (1-r) d$, i.e.,
\begin{align}\label{equa-B-H-1}
\mathcal{B}_{H-1} = \{ T_{j} \in\mathcal{T} |  \gamma(j, d) \in [\nu, H-1],  t_{j, \delta^{\prime}} < (1-r) d, t_{j, \gamma(j,d)} < r d \}.
\end{align}
{\em The second class of tasks} is defined as
\begin{align}\label{equa-second-class}
\boldsymbol{\mathcal{A}^{\prime\prime}} = \mathcal{B}_{\nu-1} \cup \mathcal{B}_{H-1}.
\end{align}
For each task $T_{j}$ with $\gamma(j,d)\leq H-1$, the relation (\ref{bound-1b}) ensures {by Property~\ref{property-deltak-tasks}} that the speedup is linear when the number of processors assigned to $T_{j}$ changes from $\gamma(j,d)$ to $\delta^{\prime}$; when assigned $\delta^{\prime}$ processors, its execution time $t_{j,\delta^{\prime}}$ satisfies
\begin{align}\label{equa-linear-bound}
t_{j,\delta^{\prime}}  = t_{j,\gamma(j,d)} \frac{\gamma(j,d)}{\delta^{\prime}}.
\end{align}

\begin{lemma}\label{lemma-B-nu-1}
For each task $T_{j}\in\mathcal{B}_{\nu-1}$, we have $t_{j,\delta^{\prime}}< (1-r) d$.
\end{lemma}
\begin{proof}
The execution time of $T_{j}$ satisfies
\begin{align*}
t_{j,\delta^{\prime}} \overset{(a)}{=} t_{j,\gamma(j,d)} \frac{\gamma(j,d)}{\delta^{\prime}} \overset{(b)}{<} \frac{\nu-1}{\delta^{\prime}}rd \overset{(c)}{<} (1-r)d
\end{align*}
where the above (a), (b) and (c) are due to Equation (\ref{equa-linear-bound}), Equation (\ref{equa-B-nu-1}) and the relation (\ref{bound-2}) respectively.
\end{proof}

\begin{proposition}\label{lemma-less-1-r}
For every task $T_{j}\in\mathcal{A}^{\prime\prime}$, we have $t_{j,\delta^{\prime}} < (1-r) d$.
\end{proposition}
\begin{proof}
It follows from Lemma~\ref{lemma-B-nu-1} and the definition of $\mathcal{B}_{H-1}$ in Equation (\ref{equa-B-H-1}).
\end{proof}

Finally, the remaining are all tasks with $\gamma(j,d)\in [\nu, H-1]$ and each has an execution time $t_{j,\gamma(j,d)}< r d$ when assigned $\gamma(j,d)$ processors and $t_{j,\delta^{\prime}}\geq (1-r) d$ when assigned $\delta^{\prime}$ processors. For each $h\in [\nu, H-1]$, {\em a single class of tasks} $\mathcal{A}_{h}$ is defined to contain all such tasks with $\gamma(j,d)=h$, {\em i.e.,}
\begin{align}\label{equa-A-h}
\boldsymbol{\mathcal{A}_{h}}= \{ T_{j}\in\mathcal{T} | \gamma(j,d)=h, t_{j,h} < r d, t_{j,\delta^{\prime}} \geq (1-r) d\}.
\end{align}


\begin{proposition}\label{lemma-x-h}
When a task is assigned $\delta^{\prime}$ processors, we have that
\begin{itemize}
\item [(\rmnum{1})] for every task $T_{j}\in\mathcal{A}_{h}$, its execution time $t_{j,\delta^{\prime}}$ is $< l_{h}d$ where $l_{h} = \frac{h}{\delta^{\prime}} r$;
\item [(\rmnum{2})] the aggregate execution time of any $x_{h}$ tasks of $\mathcal{A}_{h}$ is in $[r d, d]$.
\end{itemize}
\end{proposition}
\begin{proof}
The relation (\ref{bound-1}) implies that $\nu$ is the maximum possible integer such that the relation (\ref{bound-2}) can hold.
Let us consider every task $T_{j}\in \mathcal{A}_{h}$ and {by the definition of $\mathcal{A}_{h}$ in Equation (\ref{equa-A-h})}, we have
\begin{align}
& t_{j,\gamma(j,d)}< rd \label{equa-A-h-1} \\
& t_{j,\delta^{\prime}}\geq (1-r)d \label{equa-A-h-2}.
\end{align}
where $\gamma(j,d)=h$. We have by Lemma~\ref{lemma-monotonic-property} that the execution time of this task $T_{j}$ satisfies
\begin{align}\label{equa-gamma-h}
t_{j,\gamma(j,d)}> \frac{h-1}{h}d.
\end{align}
{Thus, by Equation (\ref{equa-linear-bound}), we have}
\begin{align}
& t_{j,\delta^{\prime}} = t_{j,h} \frac{h}{\delta^{\prime}} \overset{(d)}{<} \frac{h}{\delta^{\prime}}rd \label{equa-length-task-h} \\
& t_{j,\delta^{\prime}} = t_{j,h} \frac{h}{\delta^{\prime}} \overset{(e)}{>} \frac{(h-1)d}{h}\frac{h}{\delta^{\prime}}=\frac{h-1}{\delta^{\prime}}d  \label{equa-length-task-h-greater}
\end{align}
where the above (d) is due to Inequality (\ref{equa-A-h-1}), and (e) is due to Inequality (\ref{equa-gamma-h}). By Inequalities (\ref{equa-A-h-2}), (\ref{equa-length-task-h}) and (\ref{equa-length-task-h-greater}), we have {for any task $T_{j}\in\mathcal{A}_{h}$ that}
\begin{align*}
t_{j,\delta^{\prime}}\in \left[ \max\left\{1-r, \frac{h-1}{\delta^{\prime}}\right\}d,\, \frac{h}{\delta^{\prime}} r d \right].
\end{align*}
While executing any $x_{h}$ tasks of $\mathcal{A}_{h}$ one by one on $\delta^{\prime}$ processors, the relations (\ref{bound}) and (\ref{boundb}) ensure that their aggregate execution time is in $[r d, d]$. Together with Inequality (\ref{equa-length-task-h}), Proposition~\ref{lemma-x-h} thus holds.
\end{proof}


Proposition~\ref{lemma-less-1-r} and \ref{lemma-x-h} enable us to design good schedules. Executing as many tasks from $\mathcal{A}^{\prime\prime}$ as possible on $\delta^{\prime}$ processors by time $d$ can lead to that these processors have a utilization $\geq r$ in $[0, d]$. This also holds for the tasks of $\mathcal{A}_{h}$ where $h\in$ $[\nu, H-1]$ since at least $x_{h}$ tasks can be finished by time $d$.

\begin{proposition}\label{proposi-parameter-setting}
For a given linear-speedup threshold $\delta\geq 5$, {let $u=\left\lceil \sqrt[2]{\delta} \right\rceil-1$} where $u\geq 2$ and $\delta\in [u^{2}+1, (u+1)^{2}]$. A feasible solution that satisfies the relations~(\ref{bound-1b})-(\ref{boundb}) is as follows:
\begin{equation}\label{equa-parameters-setting}
\begin{split}
 H & = u+2 \\
 \delta^{\prime} & = u^{2}+1 \\
 \nu & = u \\
 x_{h} & = 2u+1-h \enskip \text{ for all } h\in \{\nu, H-1\}
\end{split}
\end{equation}
where $r=\frac{u+1}{u+2}$.
\end{proposition}
\begin{proof}
The proof is about verifying that the setting in Equation (\ref{equa-parameters-setting}) can satisfy the relations~(\ref{bound-1b})-(\ref{boundb}) and its detail can be found in \ref{appendix-proof-proposi-parameter-setting}.
\end{proof}

In the rest of this paper, we will set the parameter values in the way described in Proposition~\ref{proposi-parameter-setting}. Since $\nu=u$ and $H=u+2$, the tasks of $\mathcal{T}$ are finally classified as $\mathcal{A}^{\prime}, \mathcal{A}_{u+1}, \mathcal{A}_{u}, \mathcal{A}^{\prime\prime}$. Finally, we show the time complexity while classifying the tasks of $\mathcal{T}$. {When a task $T_{j}$ is allocated $p\in [1, m]$ machines, the speedup $\eta_{j,p}$ can be accessed via some oracle in constant time \citep{Jansen18a}, e.g., the oracle can obtain such information by benchmarking studies \citep{Dutton08}. Theoretically, the value of $k_{j}$ or $\delta_{j}$ is a fixed integer in $[1, m]$ and can be obtained by binary search, leading to the proposition below.}

\begin{proposition}\label{proposi-k-j-complexity}
{For each task $T_{j}\in\mathcal{T}$, the time complexity of finding the value of $k_{j}$ or $\delta_{j}$ is $\mathcal{O}(\log{m})$.}
\end{proposition}

\begin{proposition}\label{proposi-classification-complexity}
{Given the value of $d$ and the values of $k_{j}$ and $\delta_{j}$ of each task $T_{j}\in\mathcal{T}$}, the time complexity of task classification is {$\mathcal{O}(n\log{m})$}.
\end{proposition}
\begin{proof}
{The time complexity of finding the value of $\delta=\min\nolimits_{T_{j}\in\mathcal{T}}\{\delta_{j}\}$ in Equation (\ref{equa-min-delta-max-k}) is $\mathcal{O}(n)$. We can directly compute the value of $\delta^{\prime}$ by Equation (\ref{equa-parameters-setting}).} Afterwards, we classify each task $T_{j}$ where we need to check the value of $\gamma(j,d)$, $t_{j,\gamma(j,d)}$, or $t_{j,\delta^{\prime}}$ at most four times, as illustrated in Fig.~\ref{Fig-Classification}; {the time complexities of find these values determine the time complexity of classifying a task}.
{Given the execution time $t_{j,1}$ on one processor, $\gamma(j,d)=\min\{p\in [1, k_{j}]\, |\, t_{j,p}\leq d\}$ can be found by binary search with a time complexity of $\mathcal{O}(\log{k_{j}}){\leq \mathcal{O}(\log{m})}$, {where $k_j\leq m$}.} Given the values of $\gamma(j,d)$ and $\delta^{\prime}$, $t_{j,\gamma(j,d)}$ and $t_{j,\delta^{\prime}}$ can directly be computed in time $\mathcal{O}(1)$. Thus, the time complexity of classifying the $n$ tasks is {$\mathcal{O}(n\log{m})$}.
\end{proof}

\begin{algorithm*}[!b]
\DontPrintSemicolon
\SetKwInOut{Input}{Input}
\SetKwInOut{Output}{Output}
\BlankLine

\SetNoFillComment

Set the parameters by Proposition~\ref{proposi-parameter-setting} and classify the tasks of $\mathcal{T}$\;


$m^{\prime}\gets m$,\, $(\mathcal{X}^{\prime}, \mathcal{X}_{u+1}, \mathcal{X}_{u}, \mathcal{X}_{u-1})\gets (\mathcal{A}^{\prime}, \mathcal{A}_{u+1}, \mathcal{A}_{u}, \mathcal{A}^{\prime\prime})$\tcp*[f]{\footnotesize{$X^{\prime}, \mathcal{X}_{u+1}, \mathcal{X}_{u}, \mathcal{X}_{u-1}$: the currently unassigned tasks}}

\While{$\mathcal{X}^{\prime}\neq \emptyset$ and $k \leq m^{\prime}$}{

    Get an arbitrary task $T_{j}$ off $\mathcal{X}^{\prime}$: $\mathcal{X}^{\prime}\leftarrow \mathcal{X}^{\prime}-\{T_{j}\}$\;

    Assign $T_{j}$ onto $\gamma(j,d)$ idle processors: $m^{\prime}\leftarrow m^{\prime}-\gamma(j,d)$\;
}

\textbf{if} $\mathcal{X}^{\prime}\neq \emptyset$ and $m^{\prime}<k$, \textbf{then} exit\;


\textbf{if} $\bigcup\nolimits_{l^{\prime}=u-1}^{u+1}{\mathcal{X}_{l^{\prime}}} \neq \emptyset$, \textbf{then} let $l$ be the maximum integer in $\{u-1, u, u+1\}$ with $\mathcal{X}_{l}\neq \emptyset$\;


\While{$\bigcup\nolimits_{l^{\prime}=u-1}^{u+1}{\mathcal{X}_{l^{\prime}}} \neq \emptyset$ and $\delta^{\prime} \leq m^{\prime}$}{

    Get $\delta^{\prime}$ idle processors: $m^{\prime}\leftarrow m^{\prime}-\delta^{\prime}$\;

    $\mathcal{T}_{\delta^{\prime}} \leftarrow \emptyset$, $t\leftarrow 0$\tcp*[l]{\footnotesize{$\mathcal{T}_{\delta^{\prime}}$: the tasks currently chosen for the $\delta^{\prime}$ processors; $t$: the aggregate execution time of $\mathcal{T}_{\delta^{\prime}}$}}

    \While{$\mathcal{X}_{l}\neq \emptyset$}{

        Get an arbitrary task $T_{j}$ from $\mathcal{X}_{l}$\;

        \If{$t+t_{j,\delta^{\prime}}\leq d$}{

            $t\leftarrow t+t_{j,\delta^{\prime}}$,\, $\mathcal{T}_{\delta^{\prime}} \leftarrow \mathcal{T}_{\delta^{\prime}}\cup \{T_{j}\}$,\, $\mathcal{X}_{l}\leftarrow \mathcal{X}_{l}-\{T_{j}\}$\;

        }\Else{
            break\tcp*[l]{{got} enough tasks and go to line 22}
        }

        \If{$\mathcal{X}_{l}=\emptyset$ and $l>u-1$}{

            \tcp*[l]{\footnotesize{begin to assign the tasks of $\mathcal{X}_{l-1}$, $\cdots$, $\mathcal{X}_{u-1}$}}

            \If{there is an integer $\hat{l}\in [u-1, l-1]$ such that $\mathcal{X}_{\hat{l}}\neq \emptyset$}{

                Reset $l$ to the maximum such $\hat{l}$\tcp*[l]{\footnotesize{go to line 11}}
            }\Else{
                $l\leftarrow u-1$\tcp*{\footnotesize{then, $\bigcup\nolimits_{l^{\prime}=u-1}^{u+1}{\mathcal{X}_{l^{\prime}}}$ becomes empty}}
            }
        }
    }

    Assign the tasks of $\mathcal{T}_{\delta^{\prime}}$ on the $\delta^{\prime}$ idle processors\;
}

\caption{{$Sched(d)$}}
\label{UnitAlgo}
\end{algorithm*}

\subsection{Algorithm Description}
\label{sec.algo-definition}

Now, we give the scheduling algorithm $Sched$, which is presented in Algorithm~\ref{UnitAlgo}. Let $m^{\prime}$ denote the number of idle processors; initially, $m^{\prime}=m$. $\mathcal{T}$ is partitioned into $\mathcal{A}^{\prime}$, $\mathcal{A}_{u+1}, \mathcal{A}_{u}$, $\mathcal{A}^{\prime\prime}$, and these sets are also sorted and assigned in this order where the tasks in the same set are chosen in an arbitrary order. Following this order, $Sched$ assigns tasks in the following way until all tasks of $\mathcal{T}$ are assigned or there are not enough idle processors:
\begin{enumerate}
\item [(\rmnum{1})] For each unassigned task $T_{j}\in\mathcal{A}^{\prime}$, assign it onto $\gamma(j,d)$ idle processors; then, $m^{\prime}=m^{\prime}-\gamma(j,d)$ (lines 3-5).


\item [(\rmnum{2})] If $m^{\prime}\geq \delta^{\prime}$, divide the idle processors into $\lfloor \frac{m^{\prime}}{\delta^{\prime}} \rfloor$ groups, each with $\delta^{\prime}$ processors. For each group, {get unassigned tasks of $\mathcal{A}_{u+1} \cup  \mathcal{A}_{u} \cup \mathcal{A}^{\prime\prime}$} such that their aggregate execution time on $\delta^{\prime}$ processors is $\leq d$ (lines 10-21); assign these tasks onto the group of processors (line 22).
\end{enumerate}
$k$ and $\delta^{\prime}$ are given in Equations (\ref{equa-min-delta-max-k}) and (\ref{equa-parameters-setting}). Algorithm~\ref{UnitAlgo} ends (\rmnum{1}) if there are unassigned tasks but the idle processors are not enough ($m^{\prime} < k$ in line 6 or $m^{\prime} < \delta^{\prime}$ in line 8), or (\rmnum{2}) if all tasks of $\mathcal{T}$ have been assigned.

\subsubsection{Example}

\begin{figure}[!ht]
  \begin{center}
  \includegraphics[width=3.6in]{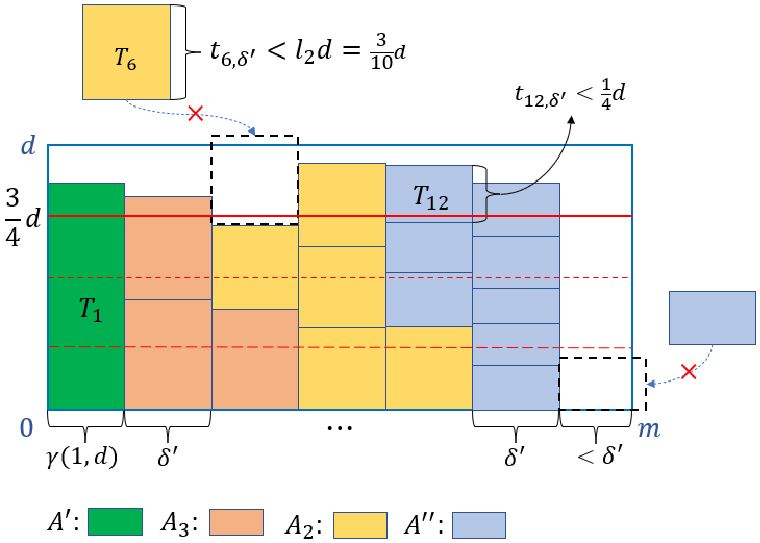}
  \end{center}
  \caption{\footnotesize{Task assignment when $\delta=5$ where each colored rectangle represents a task of some type.}}\label{Fig-Assignment}
\end{figure}


We give a toy example where $\delta=5$ to illustrate the execution of Algorithm~\ref{UnitAlgo}. By Proposition~\ref{proposi-parameter-setting}, we have $u=\nu=2$, $H=4$, $\delta^{\prime}=$ $5$, $x_{3}=2$, $x_{2}=3$, and $r=\frac{3}{4}$; then, $\mathcal{T}$ is divided into 4 subsets $\mathcal{A}^{\prime}$, $\mathcal{A}_{3}$, $\mathcal{A}_{2}$, and $\mathcal{A}^{\prime\prime}$ (line 1). Suppose that we are given $m=33$, $\mathcal{A}^{\prime}=$ $\{T_{1}\}$, $\mathcal{A}_{3}=\{T_{2}, T_{3}, T_{4}\}$, $\mathcal{A}_{2}$ $=$ $\{T_{5}, T_{6}, \cdots, T_{9}\}$, $\mathcal{A}^{\prime\prime}$ $=$ $\{T_{10}, T_{11},$ $ \cdots, T_{18}\}$ and $\gamma(1,d)=H$ for $T_{1}$. Retrospectively, we get six groups from the $m$ processors. The first group has $\gamma(1, d)$ processors, each of the remaining groups has $\delta^{\prime}$ processors, and there are also 4 ungrouped processors. As illustrated in Fig.~\ref{Fig-Assignment}, Algorithm~\ref{UnitAlgo} assigns tasks in the following way:
\begin{enumerate}
\item [(1)] Assign the only task $T_{1}$ of $\mathcal{A}^{\prime}$ onto the 1st group (lines 3-5).
\item [(2)] Assign $x_{3}=2$ tasks of $\mathcal{A}_{3}$ onto the 2nd group (lines 7-16, 22 where $l=3$).
\item [(3)] Assign the last unassigned task of $\mathcal{A}_{3}$ onto the 3rd group (lines 8-14, where $l=3$); then, $\mathcal{X}_{3}=\emptyset$ and $l$ becomes 2 (lines 17-19). Next, assign one task of $\mathcal{A}_{2}$ onto the 3rd group (lines 11-14, where $l=2$). The second task of $\mathcal{A}_{2}$ cannot be added and completed by time $d$ (lines 11-12, 15-16, 22).
\item [(4)] Assign $x_{2}=3$ tasks of $\mathcal{A}_{2}$ onto the 4th group (lines 8-16, 22 where $l=2$).
\item [(5)] Similarly to the execution of Step 3, assign the last unassigned task of $\mathcal{A}_{2}$ and three tasks of $\mathcal{A}^{\prime\prime}$ onto the 5th group (lines 8-19, 22 where $l=2,1$).
\item [(6)] Assign five tasks of $\mathcal{A}^{\prime\prime}$ onto the 6th group (lines 8-16, 22 where $l=1$).
\item [(7)] The algorithm ends when $m^{\prime}=4<\delta^{\prime}$ (line 8), although there is one unassigned task of $\mathcal{A}^{\prime\prime}$.
\end{enumerate}

By the definition of $\mathcal{A}^{\prime}$ in Equation (\ref{equa-first-class}) and Propositions \ref{lemma-less-1-r} and \ref{lemma-x-h}, the 1st-2nd and 4th-6th groups have an execution time in $[rd, d]$. The 3rd group of $\delta^{\prime}$ processors executes a mix of the tasks of $\mathcal{A}_{3}$ and $\mathcal{A}_{2}$; the aggregate execution time of tasks is $< rd$ but $\geq (1-l_{2})d$ since the rejected task of $\mathcal{A}_{2}$ has an execution time $\leq l_{2}d=3d/10$ by Proposition~\ref{lemma-x-h}. Finally, there is one unassigned task of $\mathcal{A}^{\prime\prime}$ and the number of idle processors is at most $\delta^{\prime}-1$. The total number of processors whose execution time is $<rd$ is $\delta^{\prime}+(\delta^{\prime}-1)=2\delta^{\prime}-1$. The total workload processed by the $m$ processors in $[0, d]$ is at least
\begin{align*}
w^{\prime}=(m-2\delta^{\prime}+1)rd + \delta^{\prime}(1-l_{2})d,
\end{align*}
and the overall processor utilization in $[0, d]$ is at least
\begin{equation}\label{equa-utilizatioin-example}
\begin{split}
\frac{w^{\prime}}{md} = r - \frac{r(2\delta^{\prime}-1)}{m} +  \frac{(1-l_{2})\delta^{\prime}}{m} = \frac{3}{4} - \frac{3.25}{m}.
\end{split}
\end{equation}


\subsubsection{Algorithm Analysis}



Now, we prove the features of $Sched$. 
The following conclusion is a generalization of Equation (\ref{equa-utilizatioin-example}) in the example above.

\begin{proposition}\label{UnitAlgo-Utilization}
If $Sched$ cannot schedule all tasks of $\mathcal{T}$ on the $m$ processors by time $d$, then $Sched$ achieves a processor utilization of at least
\[
\theta(\delta) = r - \frac{rk}{m}
\]
where $r=\frac{u+1}{u+2}\in (0, 1)$.
\end{proposition}
\begin{proof}
The proof is a generalization of the analysis process to derive Equation (\ref{equa-utilizatioin-example}). Please see the detailed proof in \ref{appendix-proof-proposi-unitAlgo-utilization}.
\end{proof}

\begin{proposition}\label{proposi-sched-complexity}
{Given the value of $d$ and the values of $k_{j}$ and $\delta_{j}$ of each task $T_{j}\in\mathcal{T}$,} the time complexity of Algorithm~\ref{UnitAlgo} is {$\mathcal{O}(n\log{m})$}.
\end{proposition}
\begin{proof}
The time complexity of task classification is {$\mathcal{O}(n\log{m})$} by Proposition~\ref{proposi-classification-complexity} (line 1). {Afterwards, the $n$ tasks are assigned to processors one by one} (lines 4, 12) and $Sched$ stops {when} all tasks are assigned or there are not enough processors to assign the remaining tasks, which has a time complexity of $\mathcal{O}(n)$. Hence, Algorithm~\ref{UnitAlgo} has a time complexity of {$\mathcal{O}(n\log{m})$}.
\end{proof}


Let $\mathcal{S}$ denote the tasks accepted and scheduled by Algorithm~\ref{UnitAlgo} where $\mathcal{S}\subseteq \mathcal{T}$. {$\gamma(j,d)$ denotes the minimum number of processors needed to complete $T_{j}$ by time $d$. As illustrated in Fig. \ref{Fig-Classification}, in Algorithm~\ref{UnitAlgo}, the number of processors allocated to a task is either $\gamma(j,d)$ or $\delta^{\prime}$ that is no larger than $\delta$ by Inequality (\ref{bound-1b}). By Definition \ref{def.deltak-monotonic-task} and Property~\ref{property-deltak-tasks}, we have the following lemma.}

\begin{lemma}\label{lemma-smallest}
In Algorithm~\ref{UnitAlgo}, we have for every task $T_{j}\in\mathcal{S}$ that its workload is $D_{j,\gamma(j,d)}$, which is the minimum workload needed to be processed to complete $T_{j}$ by time $d$.
\end{lemma}
\begin{proof}
Please see the detailed proof in \ref{appendix-proof-lemma-smallest}.
\end{proof}

With Proposition~\ref{UnitAlgo-Utilization} and Lemma~\ref{lemma-smallest}, we have completed the design of the scheduling algorithm $Sched$ described in Section~\ref{sec.high-idea}.

\section{Application to Two Objectives}
\label{sec.objectives}

In this section, we apply $Sched$ to respectively minimize the makespan and maximize the throughput with a common deadline $\tau$.

\subsection{Makespan Minimization}
\label{sec.makespan}

\begin{algorithm}[t]
\DontPrintSemicolon

$L\gets 0$,\, $U \gets n(\delta+2)\max_{T_{j}\in\mathcal{T}}\{t_{j,1}\}$\;

\While{$U>(1+\epsilon)L$}{

    $M\gets \frac{L+U}{2}$\;


    $\var{Flag}\gets 1$\;

    \For{$j\leftarrow 1$ \KwTo $n$}{

        \If{$\gamma(j,M)=+\infty$}{
            $\var{Flag}\gets 0$\;
            break\;
        }
    }

    \If{$\var{Flag}=1$}{
    \tcp*[l]{\footnotesize{every task $T_{j}\in\mathcal{T}$ can be completed by time $M$, i.e., $\gamma(j,d)\in [1, k_{j}]$}}

        \If{$Sched$ produces a feasible schedule of all tasks of $\mathcal{T}$ by time $M$}{
            $U\gets M$\;
        }
        \If{$Sched$ can only schedule a part of tasks of $\mathcal{T}$ by time $M$}{
            $L\gets M$\;
        }

    }\Else{
        $L\gets M$\;
    }
}

\caption{{The $OMS(\epsilon)$ algorithm}}
\label{Algo-OMS}
\end{algorithm}

Now, we give the algorithm for makespan minimization, which is formally presented in Algorithm~\ref{Algo-OMS} and also referred to as the $OMS$ algorithm (Optimized MakeSpan). Its high-level idea is as follows. Initially, let $U$ and $L$ be such that $Sched$ can produce a feasible schedule of all tasks of $\mathcal{T}$ by time $U$ but fails to do so by time $L$, {\em e.g.,} $U = n(\delta+2)\max_{T_{j}\in\mathcal{T}}\{t_{j,1}\}$ and $L=0$ (line 1); we explain the reason why such $U$ is feasible in \ref{appendix-U-initial-value}. The $OMS$ algorithm will repeatedly operate as follows and stops when $U\leq$ $(1+\epsilon)L$ (line 2):
\begin{enumerate}
\item $M\leftarrow \frac{U+L}{2}$ (line 3).
\item judge whether there exists a task $T_{j}\in\mathcal{T}$ that cannot be completed by time $M$ with the parallelism bound $k_{j}$ (lines 4-8).
\item if ${\gamma(j,M)}\in [1, k_{j}]$ for every task $T_{j}\in\mathcal{T}$ and $Sched$ can produce a feasible schedule of all tasks of $\mathcal{T}$ by time $M$, set $U\gets M$ (lines 9-11); otherwise, set $L\gets M$ (lines 9, 12-15).
\end{enumerate}

In the rest of this subsection, we analyze the approximation ratio and complexity of the algorithm.
As shown below, for a task $T_{j}$, the larger the value of $d$, the smaller the value of $\gamma(j,d)$.
\begin{lemma}\label{lemma-gamma-relation}
If $d^{\prime} < d^{\prime\prime}$ and $\gamma(j,d^{\prime}), \gamma(j,d^{\prime\prime})\in [1, k_{j}]$, then we have $\gamma(j,d^{\prime})\geq \gamma(j,d^{\prime\prime})$.
\end{lemma}
\begin{proof}
We prove this by contradiction. Suppose $\gamma(j,d^{\prime}) < \gamma(j,d^{\prime\prime})$; then we have by Property~\ref{property-deltak-tasks} that $t_{j, \gamma(j,d^{\prime})}$ $\geq t_{j, \gamma(j,d^{\prime\prime})}$. Since $t_{j, \gamma(j,d^{\prime})}\leq d^{\prime} < d^{\prime\prime}$, the minimum number of processors needed to complete $T_{j}$ by time $d^{\prime\prime}$ is no greater than $\gamma(j,d^{\prime})$, which contradicts the assumption that $\gamma(j,d^{\prime}) < \gamma(j,d^{\prime\prime})$.
\end{proof}

Let $d^{\ast}$ denote the optimal makespan. In an optimal schedule, let $D_{j}^{\ast}$ denote the workload of a task $T_{j}$ and $D^{\ast}$ denote the total workload of all tasks of $\mathcal{T}$ to be processed on the $m$ processors in $[0, d^{\ast}]$ where we have
\begin{align}\label{equa-d-star-bound}
m d^{\ast} \geq D^{\ast}.
\end{align}
When the $OMS$ algorithm ends, if $\gamma(j,L)\in [1, k_{j}]$ for every task $T_{j}\in\mathcal{T}$, only a part of tasks are scheduled by $Sched$ by time $L$ and we have by Proposition~\ref{UnitAlgo-Utilization} that $\theta(\delta)$ is a lower bound of the processor utilization in $[0, L]$; we denote by $D_{j}^{L}$ the workload of a scheduled task $T_{j}$ and by $D^{L}$ the total workload of all the scheduled tasks; here, we have
\begin{align}\label{equa-d-L-bound}
m L \geq D^{L} \geq \theta(\delta) m L.
\end{align}


\begin{lemma}\label{lemma-workload-relation}
When the $OMS$ algorithm ends, if $d^{\ast} < L$, then we have that (\rmnum{1}) $D^{\ast}\geq D^{L}$ and (\rmnum{2}) $\gamma(j,L)\in [1, k_{j}]$ for every task $T_{j}\in\mathcal{T}$.
\end{lemma}
\begin{proof}
For every $T_{j}\in\mathcal{T}$, if $d^{\ast} < L$, we have $\gamma(j,L)\in [1, k_{j}]$ since $T_{j}$ can be finished by $d^{\ast}$, with the parallelism bound $k_{j}$. By Lemma~\ref{lemma-gamma-relation}, if $d^{\prime}<d^{\prime\prime}$, we have $\gamma(j,d^{\prime})\geq \gamma(j,d^{\prime\prime})$.
Since $d^{\ast}< L$, we have in an optimal schedule that the number of processors assigned to a task $T_{j}$ is $\geq \gamma(j,d^{\ast})$, which is $\geq \gamma(j,L)$. By Property~\ref{property-deltak-tasks}, we have $D_{j}^{\ast}\geq D_{j,\gamma(j,d^{\ast})}\geq D_{j,\gamma(j,L)}$. By Lemma~\ref{lemma-smallest}, we have $D_{j,\gamma(j,L)}$ $= D_{j}^{L}$. Finally, we have $D_{j}^{\ast}\geq D_{j}^{L}$ and $D^{\ast}\geq D^{L}$.
\end{proof}

\begin{proposition}\label{theo-makespan}
  The $OMS$ algorithm gives a $\frac{1}{\theta(\delta)}(1+\epsilon)$-approximation to the makespan minimization problem with a complexity of {$\mathcal{O}(n\log{m}\log{(n m t_{m}/\epsilon)})$} {where $t_{m}=\max_{T_{j}\in\mathcal{T}}\{t_{j,1}\}$}.
\end{proposition}
\begin{proof}
For the approximation ratio, it suffices to show $U/d^{\ast}\leq \frac{1}{\theta(\delta)}(1+\epsilon)$ {where $\theta(\delta)\in (0, 1)$}. When the $OMS$ algorithm ends, we have
\begin{align}\label{equa-relation-U-L}
U\leq (1+\epsilon) L.
\end{align}
Obviously, $d^{\ast}\leq U$. In the case that $d^{*}\in [L, U]$, we have $\frac{U}{d^{\ast}} \leq 1+\epsilon \leq \frac{1}{\theta(\delta)}(1+\epsilon).$ In the other case that $d^{*}< L$, we have by Inequalities (\ref{equa-d-star-bound}), (\ref{equa-d-L-bound}) and (\ref{equa-relation-U-L}) and Lemma~\ref{lemma-workload-relation} that $$md^{*}\geq D^{\ast}\geq D^{L}\geq m\theta(\delta) L\geq m\theta(\delta) \frac{U}{1+\epsilon}.$$ Further, we have $U/d^{\ast}$ $\leq$ $(1+\epsilon)/\theta(\delta)$.

{Executing the $OMS$ algorithm needs prior knowledge of the values of $k_{j}$ and $\delta_{j}$ of all the $n$ tasks of $\mathcal{T}$, which will be used for computing the upper bound $U$ and in calling $Sched$ (lines 1, 10, 12); the time complexity of obtaining these values is $\mathcal{O}(n\log{m})$ by Proposition \ref{proposi-k-j-complexity}.} While executing the $OMS$ algorithm, the initial values of $U$ and $L$ are $n(\delta+2)t_{m}$ and 0. The binary search stops when $U\leq L(1+\epsilon)$ and the number of iterations is {$\mathcal{O}(\log{(n \delta t_{m}/\epsilon)})\leq \mathcal{O}(\log{(n m t_{m}/\epsilon)})$, where $\delta\leq m$}. At each iteration, {the time complexity of computing $\gamma(j,d)$ is $\mathcal{O}(\log{k_{j}})\leq \mathcal{O}(\log{m})$ as we show in the proof of Proposition \ref{proposi-classification-complexity}}; while judging whether there exists a task $T_{j}\in\mathcal{T}$ that cannot be completed by time $M$ (lines 4-8), the time complexity is {$\mathcal{O}(n\log{m})$}; then, $Sched$ is run (line 10 or 12) and has a time complexity {$\mathcal{O}(n\log{m})$} by Proposition \ref{proposi-sched-complexity}. The entire execution process has a time complexity {$\mathcal{O}(n\log{m}\log{(n m t_{m}/\epsilon)})$}, which is also the complexity of the $OMS$ algorithm.
\end{proof}

\subsection{Throughput Maximization with a Common Deadline}
\label{sec.model2}

Let $v_{j}^{\prime} = v_{j}/D_{j, \gamma(j, \tau)}$, and it is the maximum possible value obtained from processing a unit of workload of $T_{j}$, referred to as the (maximum) value density of $T_{j}$. We assume without loss of generality that $$v_{1}^{\prime} \ge v_{2}^{\prime} \ge \cdots \ge v_{n}^{\prime}.$$ We propose a greedy algorithm called GreedyAlgo, presented in Algorithm~\ref{GenGreedyAlgo}: it considers tasks in the non-increasing order of their value densities $v_{j}^{\prime}$ and finally finds the maximum $i^{\prime}$ such that $Sched$ can output a feasible schedule by time $\tau$ for the first $i^{\prime}$ tasks, denoted by $\mathcal{S}_{i^{\prime}}$, but fails to do so for the first $i^{\prime}+1$ tasks. The throughput of GreedyAlgo is $\sum\nolimits_{j=1}^{i^{\prime}}{v_{j}}$.

\begin{algorithm}[t]

initialize $\mathcal{S}_{i}=\{ T_{1}, T_{2},  \cdots, T_{i} \}$ for all $i\in [1, n]$\;

\For{$i\leftarrow 1$ \KwTo $n$}{

    \If{$Sched$ produces a feasible schedule of all tasks of $\mathcal{S}_{i}$ by time $\tau$}{

        $i^{\prime}\leftarrow i$\;
    }\Else{
        exit\;
    }


}
\caption{{GreedyAlgo($\tau$)}}
\label{GenGreedyAlgo}
\end{algorithm}

\begin{proposition}\label{UnitAlgo2}
GreedyAlgo gives a $\theta(\delta)$-approximation to the throughput maximization problem with a common deadline and it has a complexity of {$\mathcal{O}(n^{2}\log{m})$}.
\end{proposition}



In the rest of this subsection, we give an overview of the proof of Proposition~\ref{UnitAlgo2}. By Proposition~\ref{UnitAlgo-Utilization}, $\theta(\delta)$ is a lower bound of the processor utilization when $Sched$ schedules $\mathcal{S}_{i^{\prime}}$ in $[0, \tau]$. Let $\mathcal{OPT}$ denote the optimal throughput of our problem. The proof of Proposition~\ref{UnitAlgo2} has two parts:
\begin{itemize}
\item [(\rmnum{1})] We give an upper bound of $\mathcal{OPT}$, denoted by $\overline{\mathcal{OPT}}$, {\em i.e.,}
\begin{align}\label{equa-bound-OPT}
\overline{\mathcal{OPT}}\geq \mathcal{OPT}
\end{align}
where $\overline{\mathcal{OPT}}$ will be specified in Equation (\ref{equa-specific-bound}).
\item [(\rmnum{2})] We show that $\theta(\delta)$ is a lower bound of the ratio of the throughput of GreedyAlgo to the upper bound, {\em i.e.,}
    \begin{align}\label{equa-bound-ratio}
      \frac{\sum\nolimits_{j=1}^{i^{\prime}}{v_{j}}}{\overline{\mathcal{OPT}}} \geq \theta(\delta).
    \end{align}
\end{itemize}
Then, we have by Inequalities (\ref{equa-bound-OPT}) and (\ref{equa-bound-ratio}) that $$\frac{\sum\nolimits_{j=1}^{i^{\prime}}{v_{j}}}{\mathcal{OPT}}\geq \frac{\sum\nolimits_{j=1}^{i^{\prime}}{v_{j}}}{\overline{\mathcal{OPT}}}\geq \theta(\delta).$$ Thus, the throughput $\sum\nolimits_{j=1}^{i^{\prime}}{v_{j}}$ of GreedyAlgo is at least $\theta(\delta)$ times the optimal throughput $\mathcal{OPT}$ and GreedyAlgo is a $\theta(\delta)$-approximation algorithm.

{\em For the first part}, let us consider a fractional knapsack problem \citep{CKP} and there are a knapsack of size $\tau m$ and $n$ divisible items. With abuse of notation, each item is still denoted by $T_{j}$, with a fixed size $D_{j, \gamma(j, \tau)}$ and a value $v_{j}$. Its optimal solution is packing into the knapsack the first $\sigma$ items, denoted by $\mathcal{S}^{\prime}$, with the highest value densities such that their total size equals $\tau m$: $\sum\nolimits_{j=1}^{\sigma-1}{D_{j, \gamma(j, \tau)}} + \alpha D_{\sigma, \gamma(\sigma, \tau)}=\tau m$ where $\alpha\in (0, 1]$ and the $\sigma$-th item may be partially packed. 
The following lemma completes the description of the first part.

\begin{lemma}\label{lemma-upper-bound}
An upper bound of $\mathcal{OPT}$ is
\begin{align}\label{equa-specific-bound}
\overline{\mathcal{OPT}}=\sum\nolimits_{j=1}^{\sigma-1}{v_{j}} + \alpha v_{\sigma},
\end{align}
which is the optimal value of the knapsack problem.
\end{lemma}
\begin{proof}
GreedyAlgo chooses a subset of tasks $\mathcal{S}_{i^{\prime}} = \{T_{1}, T_{2},$ $\cdots, T_{i^{\prime}} \}$ and uses $Sched$ to schedule $\mathcal{S}_{i^{\prime}}$ on the $m$ processors in $[0, \tau]$. We will show that any solution to the problem of this paper corresponds to a feasible solution to the above knapsack problem, where the same tasks/items are chosen and the two solutions have the same total value of tasks/items; the lemma thus holds. Specifically, when a task $T_{j}\in \mathcal{S}_{i^{\prime}}$ is chosen in our problem and assigned $p_{j}$ processors, we can correspondingly pack an item $T_{j}$ with a size $D_{j, \gamma(j, \tau)}$ into the above knapsack. By Lemma~\ref{lemma-smallest}, $D_{j, p_{j}}$ $= D_{j, \gamma(j, \tau)}$ and $\sum\nolimits_{T_{j}\in \mathcal{S}_{i^{\prime}}}{D_{j, \gamma(j, \tau)}}\leq \tau m$; thus, the items $T_{1}$, $T_{2}$, $\cdots$, $T_{i^{\prime}}$ can successfully be packed into the knapsack.
\end{proof}

{\em For the second part}, the detailed proof of (\ref{equa-bound-ratio}) will be provided in \ref{appendix-proof}. Below, we provide the underlying intuition while proving (\ref{equa-bound-ratio}). The workload of each task $T_{j}\in\mathcal{S}_{i^{\prime}}$ accepted by GreedyAlgo is also $D_{j,\gamma(j,d)}$ by Lemma~\ref{lemma-smallest}. $\mathcal{S}_{i^{\prime}}$ and $\mathcal{S}^{\prime}$ contain the first $i^{\prime}$ and $\sigma$ tasks with the highest value densities respectively. We have $i^{\prime}\leq \sigma$ since in GreedyAlgo the utilization of the $m$ processors in $[0, \tau]$ is $\leq 1$. Thus, the average value density of $\mathcal{S}_{i^{\prime}}$ is no smaller than the average value density of $\mathcal{S}^{\prime}$, {\em i.e.,} $$\frac{\sum\nolimits_{j=1}^{i^{\prime}}{v_{j}}}{\sum\nolimits_{j=1}^{i^{\prime}}{D_{j,\gamma(j,d)}}} \geq \frac{\overline{\mathcal{OPT}}}{\tau m}.$$ Further, we can prove (\ref{equa-bound-ratio}): $$\frac{\sum\nolimits_{j=1}^{i^{\prime}}{v_{j}}}{\overline{\mathcal{OPT}}} \geq \frac{\sum\nolimits_{j=1}^{i^{\prime}}{D_{j,\gamma(j,d)}}}{\tau m}\geq \theta(\delta).$$ 

{Executing GreedyAlgo needs prior knowledge of the values of $k_{j}$ and $\delta_{j}$ of all the $n$ tasks of $\mathcal{T}$, which will be used in calling $Sched$ (line 3); the time complexity of obtaining these values is $\mathcal{O}(n\log{m})$ by Proposition \ref{proposi-k-j-complexity}.} During its execution, it considers $\mathcal{S}_{1}$, $\mathcal{S}_{2}$, $\cdots$, $\mathcal{S}_{n}$ one by one (line 2). Whenever $Sched$ attempts to schedule {the} tasks {of $\mathcal{S}_{i}$} on $m$ processor by time $\tau$ (line 3), it has a time complexity {$\mathcal{O}(n\log{m})$} by Proposition~\ref{proposi-sched-complexity}. Thus, the entire execution process has a time complexity {$\mathcal{O}(n^{2}\log{m})$}, which is also the complexity of GreedyAlgo.

\section{Conclusions}
\label{sec.conclusion}

In this paper, we study the problem of scheduling $n$ independent moldable tasks on $m$ processors that arises in large-scale parallel computations. For makespan minimization, the best known result is a $(\frac{3}{2}+\epsilon)$-approximation algorithm with a complexity linear in $n$ and polynomial in $\log{m}$ and $\frac{1}{\epsilon}$, where $\epsilon$ is arbitrarily small; it is achieved under a monotonic assumption: the execution time of a task $T_{j}$ is non-increasing and its workload is non-decreasing in the number $p$ of assigned processors. We propose a new perspective of the existing speedup models: the speedup of a task $T_{j}$ is linear when $p$ is small (up to a threshold $\delta_{j}$); afterwards, there may be a larger threshold $k_{j}$ such that the task is strictly monotonic when $p$ ranges in $[\delta_{j}, k_{j}]$; the bound $k_{j}$ indicates an unacceptable overhead when parallelizing the task on too many processors. Let $\delta$ be the minimum linear-speedup threshold of all tasks and $k$ be the maximum parallelism bound of all tasks. For any $\delta\geq 5$, let $u=\lceil \sqrt[2]{\delta} \rceil-1$. A main algorithmic result of this paper is a $\frac{1}{\theta(\delta)} (1+\epsilon)$-approximation algorithm for makespan minimization with a complexity {$\mathcal{O}(n\log{m}\log{(n m t_{m}/\epsilon)})$} where $\theta(\delta) = \frac{u+1}{u+2}\left( 1- \frac{k}{m} \right)$ ($m\gg k$); typically, $\delta$ can range in $[25, 150]$. As a by-product, we also propose a $\theta(\delta)$-approximation algorithm for throughput maximization with a common deadline with a complexity {$\mathcal{O}(n^{2}\log{m})$}.


\section*{Acknowledgements}

The work of Xiaohu Wu has been partially supported by the National Key R\&D Program of China (2022YFB2902900). The work of Patrick Loiseau has been partially supported by MIAI@Grenoble Alpes (ANR-19-P3IA-0003), by the French National Research Agency (ANR) through grant ANR-20-CE23-0007 and through the ``Investissements d'avenir" program (ANR-15-IDEX-02); and by the Alexander von Humboldt Foundation.


\bibliography{mybibfile-1}

\appendix

\section{S-reduction}
\label{appendix-monotonic}

For a given objective, the problem of offline scheduling of independent moldable tasks on identical machines under the $(\delta_{j}, k_{j})$-monotonic model is referred to as the problem $A$, while its counterpart under the monotonic model is referred to as the problem $B$.
let $\mathcal{OPT}_{A}$ and $\mathcal{OPT}_{B}$ denote the optimal objective function values of the two problems $A$ and $B$; here, the objective function can be either makespan minimization or throughput maximization. Let $c_{A}$ and $c_{B}$ denote the objective function values of the two problems $A$ and $B$. A S-reduction from $A$ to $B$ is formally defined as follows \citep{crescenzi1997short,crescenzi2016complexity}:
\begin{definition}
  A pair of functions $(f, g)$ is a S-reduction from $A$ to $B$ if all of the following conditions are met: (1) functions $f$ and $g$ are computable in polynomial time; (2) if $x$ is an instance of problem $A$, then $f(x)$ is an instance of problem $B$, and $\mathcal{OPT}_{B}(f(x)) = \mathcal{OPT}_A(x)$; (3) if $y$ is a solution to $f(x)$, then $g(x, y)$ is a solution to $x$, and $c_{A}(x, g(x, y)) = c_{B}(f(x), y)$.
\end{definition}

\begin{proposition}
The problem $A$ is S-reducible to the problem $B$, where $f$ and $g$ have the same time complexity $\mathcal{O}(n)$.
\end{proposition}
\begin{proof}
Let $x$ denote a specific set of $n$ independent $(\delta_{j}, k_{j})$-monotonic moldable tasks $\{T_{1}, T_{2}, \cdots, T_{n}\}$ for the problem $A$. For each task $T_{j}\in x$, its execution time is non-increasing and its workload is non-increasing in the number $p$ of processors allocated to it when $p\in [1, k_{j}]$. We also construct another task $T_{j}^{\prime}$ as follows: (\rmnum{1}) it has the same speedup feature as $T_{j}$ when $p\in [1, k_{j}]$, (\rmnum{2}) if $T_{j}^{\prime}$ is allocated more than $k_{j}$ processors (i.e., $p > k_{j}$), its execution time and workload cease to change, i.e., $t_{j,p}^{\prime}=t_{j,k_{j}}^{\prime}$ and $D_{j,p}^{\prime}=D_{j,k_{j}}^{\prime}$ where $t_{j,p}^{\prime}$ is the execution time of $T_{j}^{\prime}$ and $D_{j,p}^{\prime}$ is the workload of $T_{j}^{\prime}$ when it is allocated $p$ processors, and (\rmnum{3}) all other possible features of $T_{j}^{\prime}$ are the same as $T_{j}$, such as the execution time on one processor and the task value; here, allocating $T_{j}^{\prime}$ more than $k_{j}$ processors does not bring any benefit although there is no parallelism constraint on $T_{j}^{\prime}$. Each task $T_{j}^{\prime}$ in $f(x)$ is a monotonic task. Let $f(x)=\{T_{1}^{\prime}, T_{2}^{\prime}, \cdots, T_{n}^{\prime}\}$, which is an instance of $B$. The time complexity of constructing $f(x)$ from $x$ is $\mathcal{O}(n)$.

Suppose $y$ is an optimal or approximate solution to $f(x)$; $y$ defines a feasible schedule of $f(x)$ that determines the number $p_{j}^{\prime}$ of processors allocated to each task $T_{j}^{\prime}\in f(x)$ and the time interval $[a_{j}^{\prime}, e_{j}^{\prime}]$ in which $T_{j}^{\prime}$ is executed. Each $T_{j}$ in $x$ uniquely corresponds to a task $T_{j}^{\prime}$ in $f(x)$, and vice versa. The following function $g$ transforms the solution $y$ into a feasible solution $g(x, y)$ to $x$: for each scheduled task $T_{j}^{\prime}$ in $f(x)$ for the problem $B$, allocate $\min\{k_{j}, p_{j}^{\prime}\}$ processors to $T_{j}$ and execute $T_{j}$ in the same time interval $[a_{j}^{\prime}, e_{j}^{\prime}]$ when it comes to the problem $A$; if a task $T_{j}^{\prime}$ in $f(x)$ is not scheduled, the corresponding $T_{j}$ in $x$ is not scheduled either. Obviously, $g$ can be computed with a time complexity $\mathcal{O}(n)$. In the solutions $g(x, y)$ and $y$, $T_{j}\in x$ and $T_{j}^{\prime}\in f(x)$ have the same workload and are finished at the same time, if they are scheduled. Thus, the two solutions have the same objective function value, e.g., the same makespan or throughput. Thus, we have $c_{A}(x, g(x, y)) = c_{B}(f(x), y)$, and
\begin{align}\label{equa-left-direction}
\mathcal{OPT}_{A}(x) \geq \mathcal{OPT}_{B}(f(x)).
\end{align}

Conversely, if $y^{\prime}$ is an optimal solution to $x$ for the problem $A$ in which the number of processors allocated to each task $T_{j}\in x$ is $p_{j}$ and the time interval in which $T_{j}$ is executed is $[a_{j}, e_{j}]$. Then, this solution to $x$ is also a solution to $f(x)$ for the problem $B$. Thus, the two solutions have the same objective function value. we thus have
\begin{align}\label{equa-right-direction}
\mathcal{OPT}_{A}(x) \leq \mathcal{OPT}_{B}(f(x)).
\end{align}
By (\ref{equa-left-direction}) and (\ref{equa-right-direction}), we have $\mathcal{OPT}_{A}(x) = \mathcal{OPT}_{B}(f(x))$.
\end{proof}

\section{Proof of Proposition \ref{proposi-parameter-setting}}
\label{appendix-proof-proposi-parameter-setting}

We can easily verify that the setting in Equation (\ref{equa-parameters-setting}) satisfies the relation (\ref{bound-1b}). We have
\begin{align*}
\frac{r\nu}{\delta^{\prime}} = \frac{u+1}{u+2}\frac{u}{u^{2}+1} \overset{(a)}{\geq} \frac{1}{u+2} = 1-r
\end{align*}
where the above (a) is due to $u(u+1)\geq u^{2}+1$; thus, the relation (\ref{bound-1}) is satisfied. We have
\begin{align*}
\frac{r(\nu-1)}{\delta^{\prime}}=\frac{u+1}{u+2}\frac{u-1}{u^{2}+1}<\frac{1}{u+2}=1-r;
\end{align*}
thus, the relation (\ref{bound-2}) is satisfied.

We have $h\in [\nu, H-1] = [u, u+1]$ by Equation (\ref{equa-parameters-setting}). In the following, we first prove that the relations (\ref{bound}) and (\ref{boundb}) hold when $h=u$. We have
\begin{align}\label{bound-setting}
r\frac{u}{\delta^{\prime}} x_{u} = \frac{u+1}{u+2}\frac{u}{u^{2}+1}(u+1) \overset{(b)}{\leq} 1
\end{align}
where (b) is due to that $(u+1)^{2}u-(u+2)(u^{2}+1)=-2<0$. Thus, the relation (\ref{bound}) holds when $h=u$. We have
\begin{align}\label{equa-max-branch}
\max\left\{1-r, \frac{u-1}{\delta^{\prime}} \right\}  = \max\left\{\frac{1}{u+2}, \frac{u-1}{u^{2}+1} \right\}  \overset{(c)}{=}
\begin{cases}
 \frac{1}{u+2} & \text{ if } 2\leq u\leq 3\\
 \frac{u-1}{u^{2}+1} & \text{ if } u\geq 4
\end{cases}
\end{align}
where (c) is due to that $(u^{2}+1) - (u+2)(u-1)=3-u$. Further, if $2\leq u\leq 3$, we have
\begin{align}\label{boundb-setting-t-3}
 \frac{1}{u+2}x_{u} = \frac{u+1}{u+2}\leq 1.
\end{align}
If $u\geq 4$, we can easily verify that
\begin{align}\label{boundb-setting-t-4}
 \frac{u-1}{u^{2}+1} x_{u} = \frac{(u-1)(u+1)}{u^{2}+1} \leq 1.
\end{align}
By Inequalities (\ref{equa-max-branch}), (\ref{boundb-setting-t-3}) and (\ref{boundb-setting-t-4}), the relation (\ref{boundb}) holds when $h=u$.

Next, we prove that the relations (\ref{bound}) and (\ref{boundb}) hold when $h=u+1$.
\begin{align}
r\frac{u+1}{\delta^{\prime}}x_{u+1} = \frac{u+1}{u+2}\frac{u+1}{u^{2}+1}u \overset{(d)}{\leq} 1
\end{align}
where (d) is again due to that $(u+1)^{2}u-(u+2)(u^{2}+1)=-2<0$. Thus, the relation (\ref{bound}) holds when $h=u+1$. We have
\begin{align*}
\max\left\{ 1-r,\, \frac{u}{\delta^{\prime}} \right\} = \max\left\{ \frac{1}{u+2},\, \frac{u}{u^{2}+1} \right\} \overset{(e)}{=} \frac{u}{u^{2}+1}
\end{align*}
where (e) is due to that $(u+2)u-(u^{2}+1)=2u-1>0$. Further, we can easily verify that
\begin{align*}
\frac{u}{u^{2}+1}x_{u+1} = \frac{u^{2}}{u^{2}+1} \leq 1.
\end{align*}
Thus, the relation (\ref{boundb}) holds when $h=u+1$.

\section{Proof of Proposition \ref{UnitAlgo-Utilization}}
\label{appendix-proof-proposi-unitAlgo-utilization}

After executing $Sched$, the $m$ processors may be divided into three parts:
\begin{enumerate}
\item [(\rmnum{1})] the first part executes the tasks of $\mathcal{A}^{\prime}$ {(lines 3-5)}, {\em e.g,} the 1st group of processors in the example above,
\item [(\rmnum{2})] the second part executes the tasks of $\mathcal{A}_{H-1}$, $\cdots$, $\mathcal{A}_{u}$, $\mathcal{A}^{\prime\prime}$ {(lines 7-22)}, {\em e.g.,} the 2nd-6th groups in the example,
\item [(\rmnum{3})] the third part is idle and not assigned any task.
\end{enumerate}
$Sched$ ends with {\em two cases}: (\rmnum{1}) $m^{\prime}<k$ (line 6), or (\rmnum{2}) $m^{\prime}<\delta^{\prime}$ (line 8). Different parts exist in each case. Our analysis proceeds by showing (a) which parts of processors exist in each case and (b) the utilization of each part. $\theta(\delta)$ is a lower bound of the ratio of the total workload processed by different parts to $md$.

First, we analyze the utilizations of the three parts. The first part of processors has a utilization $\geq r$ in $[0, d]$ by the definition of $\mathcal{A}^{\prime}$. The utilization of the third part is zero. The second part can be divided into several groups, each with $\delta^{\prime}$ processors. Let $\mathcal{A}_{u-1}=\mathcal{A}^{\prime\prime}$ for ease of exposition.
For each group, we have
\begin{enumerate}
\item [\textbf{(1)}] it is assigned the tasks purely from a single set $\mathcal{A}_{h}$ where $h\in [u-1, u+1]$  (see the second, fourth and sixth groups in the example), or
\item [\textbf{(2)}] it is a mix of the tasks of multiple sets $\mathcal{A}_{h}$, $\mathcal{A}_{h-1}$, $\cdots$, $\mathcal{A}_{h^{\prime}}$ where $u+1\geq h>h^{\prime} \geq u-1$ and $h^{\prime}\in \{u, u-1\}$.
\end{enumerate}
In the former {case}, each group has an execution time $\geq rd$ by Propositions~\ref{lemma-less-1-r} and \ref{lemma-x-h}.
In the latter, there exists a task $T_{j}$ of $\mathcal{A}_{h^{\prime}}$ that cannot be completed by time $d$:
\begin{enumerate}
\item [\textbf{(2.a)}] if $h^{\prime}=u$, the group may have an execution time $<rd$ but $\geq (1 - l_{u})d$ since $t_{j,\delta^{\prime}}< l_{u}d$ by Proposition~\ref{lemma-x-h} (see the third group in the example); by Proposition \ref{proposi-parameter-setting}, the processed workload is at least
    \begin{equation}\label{equa-w}
    \begin{split}
    w   = \delta^{\prime} (1-l_{u}) d  =  (u^{2}+1)\left( 1 - \frac{u}{u^{2}+1}\frac{u+1}{u+2} \right)d  = \frac{u^{3}+u^{2}+2}{u+2}d.
    \end{split}
    \end{equation}
\item [\textbf{(2.b)}] if $h^{\prime}=u-1$, the group {has an execution time} $\geq rd$ since $t_{j,\delta^{\prime}}< (1-r)d$ (see the fifth group in the example).
\end{enumerate}
To sum up, in the second part, there are at most $\delta^{\prime}$ processors whose utilization is $<r$ in $[0, d]$ and on which the amount of processed workload is $\geq w$.

Next, we analyze which parts of processors exist. {\em In the first case}, $Sched$ ends at line 6 and the first and third parts may exist. The third part has at most $k-1$ idle processors. Thus, the average utilization of the $m$ processors is at least
\begin{align*}
r_{1} = \frac{(m-k+1) r d }{m d} = r - r\frac{k-1}{m}.
\end{align*}
{\em In the second case}, $Sched$ ends at line 8. All the three parts may exist and the third part has at most $\delta^{\prime}-1$ processors. For the second part, there are at most $\delta^{\prime}$ processors whose utilization is $<r$. Thus, the average utilization of the $m$ processors is at least
\begin{align*}
r_{2} & = \frac{w + (m-\delta^{\prime}-(\delta^{\prime}-1))rd}{md}  \overset{(a)}{=} r -\frac{1}{m}\left( \left( 2u^{2}+1 \right)\frac{u+1}{u+2} - \frac{u^{3}+u^{2}+2}{u+2} \right) \\
& = r - \frac{1}{m}\frac{u^{3} + u^{2} + u - 1}{u+2} \overset{(b)}{=} r - \frac{1}{m}\left( \delta^{\prime}r - \frac{2}{u+2} \right)
\end{align*}
where the above (a) and (b) are due to Equation (\ref{equa-w}) and Proposition~\ref{proposi-parameter-setting}. Finally, when $Sched$ ends, a lower bound of the processor utilization is $\min\{r_{1},\, r_{2}\}$, {\em i.e.,}
\begin{equation}
\begin{split}
\theta(\delta)  = r - \max\left\{ \frac{r(k-1)}{m}, \frac{1}{m}\left( \delta^{\prime}r - \frac{2}{u+2} \right) \right\}
 \overset{(c)}{\geq} r - \frac{rk}{m}
\end{split}
\end{equation}
where (c) is because $\delta^{\prime}\leq \delta\leq k$ by Inequality (\ref{bound-1b}).

\section{Proof of Lemma \ref{lemma-smallest}}
\label{appendix-proof-lemma-smallest}

$\gamma(j,d)$ is the minimum number of processors needed to complete $T_{j}$ by time $d$. By Property~\ref{property-deltak-tasks}, $D_{j,\gamma(j,d)}$ is the minimum workload needed to be processed to complete $T_{j}$ by time $d$. In Algorithm~\ref{UnitAlgo}, the number of processors used to simultaneously execute a task is either $\gamma(j,d)$ for $\mathcal{A}^{\prime}$ or no more than $\delta$ for $\mathcal{A}_{H-1}$, $\cdots$, $\mathcal{A}_{u}$, and $\mathcal{A}^{\prime\prime}$. For the latter, by Inequality (\ref{bound-1b}), we have for each task $T_{j}$ that $\gamma(j,d)\leq$ $H-1\leq \delta^{\prime}\leq \delta$; by Property~\ref{property-deltak-tasks}, the workload of $T_{j}$ keeps constant when the number of assigned processors varies in $[0, \delta]$ and we have in Algorithm~\ref{UnitAlgo} that the workload of $T_{j}$ equals $D_{j,\gamma(j,d)}$. Thus, the lemma holds.

\section{The Initial Value of $U$}
\label{appendix-U-initial-value}

The initial value of $U$ is set as
$$U = n(\delta+2)\max\nolimits_{T_{j}\in\mathcal{T}}\{t_{j,1}\},$$ which is at least $\delta+2$ times the total execution time of all tasks when every task is assigned one processor. $\gamma(j,U)$ is the minimum number of processors needed to complete $T_{j}$ by time $U$, and we have $\gamma(j,U)=1$ for all $T_{j}\in\mathcal{T}$. We have by Inequality (\ref{bound-1b}) that $1\leq H-1\leq \delta^{\prime}\leq \delta$. By Property~\ref{property-deltak-tasks}, we have for every task $T_{j}\in\mathcal{T}$ that
\begin{align*}
t_{j,\delta^{\prime}}=\frac{t_{j,1}}{\delta^{\prime}} \leq \frac{U}{n(\delta+2)\delta^{\prime}}\leq \frac{U}{\delta+2}<\frac{U}{H} = (1-r)U
\end{align*}
where $n\geq 1$ and $r=\frac{H-1}{H}$. Every task of $\mathcal{T}$ has an execution time $< (1-r)U$ when assigned $\delta^{\prime}$ processors. Thus, all tasks of $\mathcal{T}$ are in the class $\mathcal{A}^{\prime\prime}$, and the other classes $\mathcal{A}^{\prime}$, $\mathcal{A}_{H-1}$, $\cdots$, $\mathcal{A}_{\nu}$ are empty. Now, we show that $Sched$ can produce a feasible schedule for all tasks of $\mathcal{T}$ by time $U$. All tasks of $\mathcal{T}$ constitute $\mathcal{A}^{\prime\prime}$ and will be executed one by one on $\delta^{\prime}$ processors (see lines 7-22 of Algorithm~\ref{UnitAlgo}); the total execution time of $\mathcal{T}$ is $\leq n\max\nolimits_{T_{j}\in\mathcal{T}}\{t_{j,1}\}\leq U$.

\section{The Detailed Proof of the Second Part}
\label{appendix-proof}



Below, we formally prove Inequality (\ref{equa-bound-ratio}). GreedyAlgo accepts the first $i^{\prime}$ tasks with the highest value densities $v_{j}^{\prime}$, and the achieved throughput is $\sum\nolimits_{j=1}^{i^{\prime}}{v_{j}}$. $Sched$ is used to schedule the $i^{\prime}$ tasks, and each accepted task $T_{j}$ has a workload $D_{j,\gamma(j,\tau)}$ by Lemma~\ref{lemma-smallest}. We denote by $\omega\in [0,1]$ the actual utilization of the $m$ processors in $[0, \tau]$ achieved by GreedyAlgo, {\em i.e.,}
\begin{align*}
\sum\limits_{j=1}^{i^{\prime}}{D_{j,\gamma(j,\tau)}}=\omega  \tau  m.
\end{align*}
By Proposition~\ref{UnitAlgo-Utilization}, $\theta(\delta)$ is a lower bound of the processor utilization and we have
\begin{align}\label{equa-lower-bound-utilization}
\omega \geq \theta(\delta)\in (0, 1).
\end{align}
Since $\omega\leq 1$, we have
\begin{align}\label{equa-bound-i-prime}
i^{\prime}\leq \sigma.
\end{align}

\begin{lemma}\label{monotonic}
The throughput $\sum\limits_{j=1}^{i^{\prime}}{v_{j}}$ achieved by GreedyAlgo is at least $\theta(\delta) \overline{\mathcal{OPT}}$ where $\overline{\mathcal{OPT}}$ is given in Equation (\ref{equa-specific-bound}).
\end{lemma}
\begin{proof}
By Inequality (\ref{equa-bound-i-prime}), we will analyze two cases that $i^{\prime}=\sigma$ and $i^{\prime}<\sigma$ respectively. {\em In the case that $i^{\prime}=\sigma$}, we have
\begin{align*}
\sum\limits_{j=1}^{i^{\prime}}{v_{j}} \leq \overline{\mathcal{OPT}} = \sum\limits_{j=1}^{\sigma-1}{v_{j}} + \alpha  v_{\sigma}
\end{align*}
by Lemma~\ref{lemma-upper-bound}. Thus, we have $\alpha=1$ and the lemma holds.

{\em In the case that $i^{\prime}<\sigma$}, let
\begin{align}
X_{1} & =  \frac{1}{\omega}  \sum\limits_{j=1}^{i^{\prime}}{ D_{j, \gamma(j, \tau)} } - \sum\limits_{j=1}^{i^{\prime}}{D_{j, \gamma(j, \tau)}}\nonumber\\
X_{2} & =
\begin{cases}
& \sum\limits_{j=i^{\prime}+1}^{\sigma-1}{D_{j, \gamma(j, \tau)}} + \alpha  D_{\sigma, \gamma(j, \tau)}\enskip\enskip\enskip\, \text{ if } i^{\prime}<\sigma-1\\
& \alpha  D_{\sigma, \gamma(j, \tau)}\enskip\enskip\enskip\enskip\enskip\enskip\enskip\enskip\enskip\enskip\enskip\enskip\enskip\enskip\enskip\enskip\enskip\enskip \text{ if } i^{\prime}=\sigma-1
\end{cases}\nonumber\\
Y & =
\begin{cases}
& \sum\limits_{j=i^{\prime}+1}^{\sigma-1}{v_{j}} + \alpha  v_{\sigma}\enskip\enskip\enskip\, \text{ if } i^{\prime}<\sigma-1\\
& \alpha  v_{\sigma}\,\enskip\enskip\enskip\enskip\enskip\enskip\enskip\enskip\enskip\enskip\enskip\enskip\, \text{ if } i^{\prime}=\sigma-1
\end{cases}\nonumber
\end{align}
Recall
\begin{align*}
\tau m = \frac{1}{\omega}\sum\limits_{j=1}^{i^{\prime}}{D_{j, \gamma(j, \tau)}} = \sum\limits_{j=1}^{\sigma-1}{D_{j, \gamma(j, \tau)}} + \alpha D_{\sigma, \gamma(\sigma, \tau)}.
\end{align*}
We thus have $X_{1}=X_{2}$ since $$\tau  m - X_{1} = \sum\limits_{j=1}^{i^{\prime}}{D_{j, \gamma(j, \tau)}} = \tau  m - X_{2}.$$ The total value obtained by GreedyAlgo is $\sum\limits_{j=1}^{i^{\prime}}{v_{j}}$ and we have
\begin{equation}\label{inequality-1}
\begin{split}
\frac{\sum\limits_{j=1}^{i^{\prime}}{v_{j}}}{\omega  \tau  m} & \overset{(a)}{=} \frac{ \sum\limits_{j=1}^{i^{\prime}}{v_{j}^{\prime}  \left(\frac{1}{\omega}  D_{j, \gamma(j,\tau)} - D_{j, \gamma(j, \tau)} + D_{j, \gamma(j, \tau)}\right)} }{ \tau  m}\\ & \overset{(b)}{\geq} \frac{\sum\limits_{j=1}^{i^{\prime}}{v_{j}} +  v_{i^{\prime}}^{\prime}  X_{1} }{\tau  m} \overset{(c)}{=}  \frac{\sum\limits_{j=1}^{i^{\prime}}{v_{j}} +  v_{i^{\prime}}^{\prime}  X_{2} }{\tau  m}\\ & \overset{(d)}{\geq} \frac{\sum\limits_{j=1}^{i^{\prime}}{v_{j}} + Y }{\tau  m} = \frac{\overline{\mathcal{OPT}}}{\tau  m}.
\end{split}
\end{equation}
Here, in Equation (a), $v_{j}=v_{j}^{\prime}  D_{j,\gamma(j,\tau)}$; Inequalities (b) and (d) are due to that $v_{1}^{\prime}\geq  \cdots \geq v_{i^{\prime}}^{\prime}\geq \cdots\geq v_{n}^{\prime}$; Equation (c) is due to that $X_{1} = X_{2}$. Due to Inequality (\ref{inequality-1}), we have $\sum\limits_{j=1}^{i^{\prime}}{v_{j}}\geq \omega  \overline{\mathcal{OPT}}$; further, by Inequality (\ref{equa-lower-bound-utilization}), the lemma holds.
\end{proof}

\end{document}